\documentclass[aps,pra,notitlepage,nofootinbib,longbibliography]{revtex4-1}

\usepackage{amsmath}
\usepackage{amsfonts}
\usepackage{amssymb}
\usepackage{amsthm}
\usepackage{mathtools}
\usepackage{appendix}
\usepackage[mathscr]{eucal}
\usepackage{graphicx}
\usepackage{float}
\usepackage{url}
\newcommand{\abs}[1]{\bigl\lvert#1\bigr\rvert}

\newcommand{\me}{\mathrm{e}}
\newcommand{\mi}{i} 
\newcommand{\mdiff}{d} 

\newcommand{\Int}{\int_{-\infty}^{\infty}}

\newcommand{\trace}{\mathrm{tr\,}}

\newcommand{\ie}{\emph{i.e.}}

\newcommand{\eg}{\emph{e.g.}}
\newcommand{\Eg}{\emph{E.g.}}
\newcommand{\viz}{\emph{viz.}}

\newcommand{\rhs}{\textit{rhs}}
\newcommand{\dprime}{\prime\prime}
\newcommand{\schrdngr}{Schr\"{o}dinger\,}
\newcommand{\schrdngrs}{Schr\"{o}dinger's\,} 

\newcommand{\op}{\ensuremath{\mathrm{op}}}

\newcommand{\xop}{\ensuremath{x_\mathrm{op}}}

\newcommand{\kop}{\ensuremath{k_\mathrm{op}}}
\newcommand{\pop}{\ensuremath{p_\mathrm{op}}}
\newcommand{\Hop}{\ensuremath{H_\mathrm{op}}}
\newcommand{\Vop}{\ensuremath{V_{\op}}}
\newcommand{\hop}{\ensuremath{H_0^\mathrm{op}}}

\newcommand{\pw}{\ensuremath{\mathrm{pw}}}
\newcommand{\hc}{\ensuremath{\hat{c}}}
\newcommand{\hC}{\ensuremath{\hat{C}}}
\newcommand{\hcd}{\ensuremath{\hat{c}^\dagger}}
\newcommand{\hCd}{\ensuremath{\hat{C}^\dagger}}
\newcommand{\hpsi}{\ensuremath{\hat{\psi}}}
\newcommand{\hpsid}{\ensuremath{\hat{\psi}^\dagger}}
\newcommand{\hPsi}{\ensuremath{\hat{\Psi}}}
\newcommand{\hPsid}{\ensuremath{\hat{\Psi}^\dagger}}

\newcommand{\coh}{\ensuremath{\mathrm{coh}}}

\newcommand{\wap}{\ensuremath{\mathrm{wp}}}
\newcommand{\Wp}{\ensuremath{\mathrm{wp}}}
\newcommand{\meas}{\ensuremath{\mathrm{meas}}}
\newcommand{\inst}{\ensuremath{\mathrm{inst}}}
\newcommand{\pure}{\ensuremath{\mathrm{pure}}}
\newcommand{\mix}{\ensuremath{\mathrm{mix}}}
\newcommand{\kbar}{\ensuremath{\overline{k}}}
\newcommand{\Dk}{\ensuremath{\Delta k}}

\newcommand{\kp}{\ensuremath{k^\prime}}

\newcommand{\st}{\ensuremath{\mathrm{st}}}
\newcommand{\inci}{\ensuremath{\mathrm{in}}}
\newcommand{\rf}{\ensuremath{\mathrm{rf}}}
\newcommand{\barr}{\ensuremath{\mathrm{bar}}}
\newcommand{\tr}{\ensuremath{\mathrm{tr}}}
\newcommand{\pco}[1]{\ensuremath{p_\coh(#1)}}
\newcommand{\Pco}[1]{\ensuremath{P_\coh(#1)}}
\newcommand{\gvn}{\ensuremath{|\kbar,\Delta k}}

\newcommand{\km}{\ensuremath{k_\meas}}
\newcommand{\Delkwp}{\ensuremath{\Delta k_\Wp}}

\newcommand{\Delkinst}{\ensuremath{\Delta k_\inst}}
\newcommand{\dtilk}{\ensuremath{d\tilde{k}}}

\newcommand{\xp}{\ensuremath{x^\prime}}

\newcommand{\tp}{\ensuremath{t^\prime}}

\newcommand{\zp}{\ensuremath{z^\prime}}

\newcommand{\Dbra}[1]{\langle#1\bigr|}
\newcommand{\Dket}[1]{\bigr|#1\rangle}
\newcommand{\Dbraket}[2]{\langle#1\bigr|#2\rangle}
\newcommand{\Dketbra}[2]{|#1\rangle\langle#2|}
\newcommand{\Dme}[3]{\Dbra{#1}#2\Dket{#3}}
\newcommand{\Dop}[3]{\Dket{#1}#2\Dbra{#3}}

\newcommand{\DketV}{\ensuremath{\Dket{\tilde{0}}}}

\newcommand{\mb}[1]{\mathbf{#1}} 

\newcommand{\mbr}{\mb{r}} 

\newcommand{\rgn}[1]{\ensuremath{\mathtt{#1}}}

\newcommand{\vsp}{\vspace*{10 pt}}

\newcommand{\frend}{\hfill$\Box$}

\newcommand{\FTI}{\ensuremath{\mathrm{FT}^{-1}}}

\newcounter{remark} 

\DeclareMathOperator{\sinc}{sinc}

\begin{document}
\title{On Scattering of 1-D Wave Packets}
\author{N.\ F.\ Berk}
\thanks{Retired, with research affiliations at NIST and UMD}
\affiliation{
NIST Center for Neutron Research (NCNR), National Institute of Standards and Technology (NIST), Gaithersburg, MD}
\email{norman.berk@nist.gov}
\affiliation{Department of Materials Science and Engineering (DMSE), University of Maryland (UMD), College Park, MD}

\begin{abstract}
We present an exact solution to the one--dimensional (1--D) scattering--from--a--barrier problem for an incident neutron described by a wave packet. As an aid to presenting our approach, we spend some time on a basic review of how wave packets appear in standard quantum mechanics (SQM), paying attention to interpretive aspects of a working theory both familiar on the one hand while historically subject to debates, confusion, and misunderstandings on the other. Several appendices also are included to address various mathematical issues that may be helpful to some readers.
\end{abstract}

\maketitle
\section{Introduction}\label{0DWP}

The main object of this article is to obtain an exact and computationally accessible solution to the one--dimensional (1--D) scattering--from--a--barrier problem when an incident neutron is described by a wave packet rather than by the usual (in the textbook sense) plane wave. An incident wave packet (as we use the term) is the solution of the time--dependent \schrdngr equation for the scattering problem, taken to a sufficiently early time that it can be sensibly treated as independent of the scattering barrier and thus representing a ``free particle'' state directed toward the barrier, consistent with the picture of an appropriate neutron source. The solution of this problem entails not only some mathematical development but also involves an understanding of what to do---in a manner of speaking---with the found wave packet once in hand, including what is actually measurable in terms of how we normally interpret scattering results. And such understanding, itself, must make use of carefully defined concepts that may be overlooked or misconstrued in conventionally applied scattering theory under the assumptions of plane--wave scattering. In this setting it would be out of place, of course, to attempt a comprehensive review of the vast subject of quantum mechanics; but we believe that some discussion of the basic mathematical and interpretive elements bearing on the scattering problem would be helpful. In what follows, therefore, we make an effort to be as clear as possible about what we are doing and why we are doing it.

\subsection{States, quanta, and ``particles''}\label{PQ}
The semantics of what follows is based on our viewpoint of what we will call standard quantum mechanics (SQM), \viz, the familiar mathematical prescriptions and interpretive elements found in widely used textbooks and monographs. The physical implications of some of these elements, however, are not universally agreed upon by authors and practitioners, even after a century of discussion and debate, providing opportunities for misunderstandings flowing from differing intentions in the use of a common terminology. We therefore try to present a clear picture of what we believe to be a consistent approach to the application of SQM to the problem of wave packet scattering. To this end, the Introduction reviews some basic quantum formalism with emphasis on perspectives that we consider important for interpreting neutron scattering experiments in general and, particularly, in the stated context. Section \ref{1DWP} develops in detail the special case of wave packet scattering in 1-D, a problem that has been addressed in the literature, as cited in Sec.\ \ref{1DWP}, but which continues, we believe, to raise fundamental issues.

We also include five appendices for extended mathematical material and, in some cases, to broaden the scope of the discussion for interested readers. Appendix A reviews the standard model of a Gaussian wave packet; no surprises, but some helpful detail. Appendix B derives a representation of wave packets using the structure of the Feynman integral, an approach that lends itself to building more varied wave packet models for realistic problems. Appendix C discusses wave packets from the standpoint of second quantization, a rarely treated topic at this level. Appendix D shows a mathematical property of time--dependent statistical operators (density matrices) that has been invoked in some discussions of wave packet scattering. Appendix E presents a rather complete review of the solution of the 1-D plane wave scattering problem using the powerful transfer matrix technique applied in Sec.\ \ref{1DWP}. For other applications to neutron reflectivity see, for example, \cite{MB10} and \cite{MB11} and additional references therein. Appendix E, however, offers a more thorough mathematical development than is usually found. And since we make frequent use of Dirac ``bra--ket'' methodology, Appendix F gives a brief tutorial for readers who may not be entirely comfortable with it.

We begin our description of the concepts of concern by sharply separating the notions (generally expressed as ``wave--particle duality'') of the two fundamental quantum entities: \textsl{particles}\footnote{Most treatments of SQM speak of particles in a generic sense, taking for granted the classical notions that are readily visualized. (See the last paragraph of this section for a brief mention of some relevant concepts.) Our usage of ``particles'' in what follows will always stand for quantum particles or quanta.} (or \textsl{quanta}) and \textsl{states}. Let it be enough right now to say simply that quanta are what quantum states are about---and that quantum states are represented by physically acceptable solutions of the \schrdngr equation, assigned to a Hilbert space of continuous space--time functions, or to an associated vector space projectable onto function space in the manner of the Dirac formalism. We mainly deal here with what could loosely be called ``procedural'' notions, some of which, however, unavoidably touch on philosophical topics that we tread upon only lightly, relying on sources that have been helpful to our way of thinking about them. And there are some deep philosophical issues we choose to avoid, such as the meaning of \textsl{reality}. Outside of some commonly invoked ``physical'' contexts, we do not know when to consider particles and states, quantum or otherwise, as ``real things'' beyond their mathematical representations, so we refrain from relying on such categorization.

An effective summary of our basic point of view on SQM is found in the very first sentence in Chapter 1 of Schweber's textbook \cite{Schweber}: ``Quantum Mechanics, as usually formulated, is based on the postulate that all physically relevant information about a physical system at a given instant of time is derivable from the knowledge of the state function of the system.'' Schweber at this early point does not define ``physical system,'' but it subsequently becomes clear that his ``system'' refers to a collection of quantum particles, including the singleton. More generally, in the vast quantum literature, it is also common to see references to ``system of interest,'' which may, in various contexts, also include classical elements pertaining to a ``sample,'' ``apparatus,'' or ``measurement device''---or even an ``observer.'' In this regard, we mention that \textsl{relational quantum mechanics} (RQM) is an interpretation of (otherwise) SQM that rejects the physical meaning of states of ``isolated'' systems, even as defined by solutions of the \schrdngr equation. In RQM, states are deemed physical only in relation to one another and are ultimately definable only by the measurement results they produce. See \cite{Laudisa} for a recent description of RQM. In our representation of SQM we overlook a number of such alternative viewpoints, although certain aspects of the picture we present are not necessarily at odds with them. For the purposes at hand, however, we do assume, along with virtually all SQM textbooks, that it is meaningful to consider the application of SQM to ``single--particle'' systems, as in most treatments of scattering theory.

The distinction between quantum particle and state appropriate to SQM is not to be taken for granted. A (more or less) spatially compact wave packet (\ie, a physically acceptable solution of the \schrdngr equation) ``moving along'' a certain direction (\ie, being temporally displaced in its coordinate system) with some given (group) velocity is a \textit{state} limiting the properties of a quantum particle in that state that may be revealed only by appropriate measurements. Looks aside, the functional image of the wave packet (the state) is not a ``smeared out particle.''\footnote{A reminder of a bit of history relating to this picture: In the last of his four famous 1926 papers, \cite{shrdPR1926} (published in English), \schrdngr, in Sec.\ 8, interprets the wave function according to his ``hypothesis'' that ``...the charge of the electron is not concentrated in a point, but is spread out through the whole space, proportional to [$|\psi(\mbr)|^2$]'', a view he eventually recants in light of the three papers by Max Born on the statistical interpretation of the wave function---also in 1926---that becomes known as the Born rule. (An English translation of the first of the Born rule papers, ``On The Quantum Mechanics of Collisions,'' appears in \cite{WZ}). His adoption of the Born rule might be said to lack some enthusiasm, however; \eg, in his 1935 ``cat paradox'' paper \cite{catpap}, Sec.\ 7---\textsl{The $\psi$--function as Expectation--catalog}---starts with: ``Continuing to expound on the \textsl{official teaching} [italics added]...the $\psi$--function...is now the means for predicting probability of measurement results.'' Indeed, in a 1946 letter to Einstein, \schrdngr admits that: ``God knows I am no friend of of the probability theory, I have hated it from the first moment when our dear friend Max Born gave it birth.'' See \cite{Moore}, p.\ 435. Also see \cite{Scott}, Chap. IV, for a concise review of \schrdngrs evolving interpretation of QM.}

States follow deterministic trajectories in Hilbert space (namely, the rigorous evolution of functions representing unique solutions of the \schrdngr equation) but not their quanta.\footnote{We do not consider the trajectories defined by Bohmian mechanics, which, with its added interpretation of the solutions of the \schrdngr equation, lies outside SQM. See \cite{Goldstein}, for example.} And a quantum particle is not otherwise to be confused with a billiard ball; as discussed in Sec.\ \ref{PQSu}, it is a conceptual precursor of properties produced by measurements, which may or may not to be consistent with classical ``particle--like'' behavior.

Basic attributes normally associated with the common notion of particles include \textsl{individuality} (this one, not that one), and \textsl{distinguishability} (this one is ``1'', that one, ``2''). These qualities contribute to what is known to philosophers as primitive (\ie, underivable) ``thisness,'' qualities that are problematic in the quantum domain, especially in quantum field theory (QFT). See Teller \cite{Teller1}, Ch.\ 2, for example, and \cite{Teller3}.\footnote{The formal philosophical term for thisness is \textsl{haecceity} (pronounced $hak\cdot \underline{see}\cdot ity$), taken from the Latin \textsl{haecceitas}, literally ``the quality of being this.'' The study of haecceity deals with the notion of uniqueness; how do we tell that two objects actually are two objects, not the same object as might be viewed from different perspectives? Our everyday experiences usually make this an easy decision; but in the realm of quantum physics, such experiences are not available; instead, special rules have been invented for counting and labeling ``like--objects'' and these affect how quantum states are to be constructed.} While abstract distinctions between the notions of thisness and non--thisness (or ``sameness'', as used by Schr\"{o}dinger in \cite{ESUbeity}) are mainly of concern in QFT, which deals \textsl{ab initio} with ``many--particle'' states, they play conceptual and mathematical roles in SQM as well, especially in regard to the physical meaning of superpositions, to which we now turn.

\subsection{States, quanta, and superpositions}\label{PQSu}
We take \textit{superposition} of states to mean a \textit{coherent}---\ie, simply additive---combination of solutions of the \schrdngr equation at a specified value of time. Thus, a (coherent) superposition also satisfies the \schrdngr equation. In wider contexts such superpositions are called ``pure'' states. There also exist linear forms for combining state--defined  projection operators into \textsl{statistical operators} to describe so--called ``impure'' or ``mixed'' states, thereby suppressing coherency to produce a more classical sense of state. Mixed states typically describe beams of independently created particles by appropriate sources in scattering setups.\footnote{But, to borrow from Maudlin \cite{Maudlin}, ``[there is] no possible physical process (either \schrdngr evolution or collapse) that can take [a single--particle] pure state [into] a mixed state.'' Thus ``there is no possible way to take a single particle and prepare it in such a state.''} The mathematical and conceptual connections between superposition and coherence can be slippery, however. For example, the concept of \textit{superselection}, introduced by Wick, Wightman and Wigner \cite{Wick}, asserts the existence of states having the property that pure--state superpositions behave as mixed--state (statistical operator) superpositions---or equivalently (it can be argued) that pure--state superpositions of such states simply are forbidden. Superselection generally is believed to apply to the class--defining attributes of states of elementary particles, such as rest mass, electric charge, and so on. So--called ``soft'' or ``weak'' superselection has been associated in some circles with the dynamical process called \textsl{decoherence}. A comprehensive, although mathematically advanced, ``philosophical'' view of superselection is given by \cite{Earman}. Near the end of Sec.\ \ref{1DWP} we will employ incoherent superpositions in the form of classical averaging of intensities ``on top of'' wave packet generated coherent superpositions in the problem of wave packet scattering.\footnote{To quote from Auff\`ees and Grangier \cite{AG}: ``We adopt here the usual view that statistical mixtures correspond to an extra layer of classical probabilities added over a truly quantum structure provided by pure states.''}

Adopting recommendations made by Teller in \cite{Teller1}, we view superpositions of states as \textsl{propensities} for revealing the state of affairs when---and only when---a \textsl{measurement} of the state of affairs is made. Teller's sense of propensity (or tendency) is motivated by the Born rule for associating probabilities of measurement outcomes with the (absolute squared) amplitudes of a superposition. He emphasizes the view that measurements in the quantum sense are not spontaneous (or hidden) occurrences but require what he calls purposeful ``activating conditions.'' Moreover, in Teller's view, a measurement result does not reveal a prepossessed value of the property of the quantum particle being measured, excepting its class--defining properties mentioned in the previous paragraph.\footnote{A prepossessed value of a quantum property generally is taken to mean an ``existing'' value that is unknown prior to measurement. The notion of un--prepossessed values is not original with Teller, of course. For example, in the ``cat paradox'' paper \cite{catpap}, \schrdngr writes [Sec.\ 8, p. 329]: ``The rejection of realism has logical consequences. In general, a variable \textsl{has} no definite value before I measure it; then measuring it does \textsl{not} mean ascertaining the value it \textsl{has}.''}

The contrary assertion---that a measured value (say of position or momentum) is prepossessed---has had its champions over the decades since the introduction of quantum mechanics, notably Einstein, of course. After all, one may ask: if the measured value isn't prepossessed, then what, exactly, is it the value \textsl{of}? One answer is that quanta behave according to the state they are in; and it is the \textsl{state} that at a given time ``prepossesses'' all potential values that the activating conditions of the measurement device can reveal at that time. The quantum particle, as the ``thing'' that activates the measurement device to produce a value, thus could be seen--- via the application of repeated measurements on (assumed) identically prepared states---as a tool for elucidating, in statistical fashion defined by the Born rule, certain aspects of its pre--measurement state.

One could quibble that if the state is an eigenstate of the property of concern, then surely that property is prepossessed by the quantum in that state. This view, however, would seem to ``prepossess'' knowledge of what the state is, removing any interest in measuring it. SQM dictates that any (assumed perfect) measurement of a property must reveal an eigenvalue of the operator associated with the property; this does not imply that the state itself was an eigenstate of that operator prior to measurement. SQM does require, however, that the post--measurement state be the appropriate eigenstate. This measurement--induced change of state is called \textsl{state reduction}, following von Neumann\footnote{In \cite{vNeu}}---or much more commonly, state ``collapse.'' An underlying mechanism for collapse, which is not a property of the \schrdngr equation, remains an open topic of quantum theory, which we do not pursue here.

There is, of course, the matter of what physical processes constitute measurements. In practice, certainly in scattering experiments, a measurement in the quantum sense generally is revealed by a ``click'' (or equivalent signal) of a suitable particle detector. The click occurs at some location of a detector element (or ``pointer'') and thus is essentially a position measurement at some clock--time $t$. Typically, for a given apparatus, these position values are converted to some other variable, such as an angle, and---with certain assumptions depending on the setup---may be further converted to a number representing a dynamical property of a scattered particle, such as energy or momentum. The ``intensity'' of clicks observed for an incident beam at various detector positions then can be statistically related to structural properties of the scattering sample with the aid of Born's rule and appropriate predictive theory.\footnote{We learn in our SQM courses that the only ``observable'' quantum values are eigenvalues of Hermitian operators, but in the laboratory we advertise that we have ``measured,'' say, the neutron reflectivity of some material sample. There is, of course, no Hermitian ``reflectivity'' operator as such. In the quantum realm the intended meanings of words such as \textsl{observe} and \textsl{measure} obviously can be strongly contextual.}

\subsection{States, quanta, and Fock space}\label{PQFS}

Before moving on to the specific wave packet problem of concern here, we believe it is helpful to review how superposition works for the familiar double--slit problem by placing suitable quanta into a Fock space (also called ``occupation--number space''). Fock spaces normally are employed in the many--particle problem, but they also can be put to use in the ``many--state'' problem when presented with multiple choices for single--quantum states. Thus, in the usual textbook 2--dimensional $(x,z)$ picture, let us consider one quantum (the ``source'' quantum) propagating along the $x$--direction and incident at $x=0$, say, upon a screen having two slits oriented along the $z$--axis and with a detector screen along $z$, some distance away, at $x=D>>0$. First we define Fock states for two slits: let $\Dket{n_1,n_2}$ be a state associated with the possible passage of $n_1$ quanta through slit 1 and $n_2$ quanta through slit 2. Since we are dealing here with an assumed single source quantum (a fermion), the only such states of concern can be $\Dket{\phi_1}=\Dket{1,0}$ and $\Dket{\phi_2}=\Dket{0,1}$, associated with propensities relating to the possibilities that the incident quantum passes through slit 1 but not slit 2, or through slit 2 but not slit 1. We can then  define the hermitian number operators $N_{\op,i}$ to ``measure'' the number of quanta associated with each Fock state, here either 1 or 0. Then the state $\Dket{\phi_i}$ is an eigenstate of $N_{\op,i}$ belonging to the eigenvalue $1$; \ie, $N_{\op,i}\Dket{\phi_i}=1\times\Dket{\phi_i}$, for $i=1,2$, while $N_{\op,i}\Dket{\phi_j}=0\times\Dket{\phi_j}=0$ for $i\ne j$. These $\Dket{\phi_i}$ can  be taken to be normalized to unity, and, as defined, are easily shown to be orthogonal to one another: for $j\neq i$,

\begin{equation}\label{eqPQ30}
\begin{split}
\Dbraket{\phi_j}{\phi_i}&=(N_{\op,j}\Dket{\phi_j})^\dagger(N_{\op,i}\Dket{\phi_i}) = \Dme{\phi_j}{N_{\op,j}N_{\op,i}}{\phi_i}\\
&=\Dbra{\phi_j}\bigl(N_{\op,j}(1\times\Dket{\phi_i})\bigr)=\Dbra{\phi_j}(0\times\Dket{\phi_i})=0\,.
\end{split}
\end{equation}
Furthermore, let us assume---absent any other knowledge of the situation---that the passage of the source quantum through the slit screen ``prepares'' the linear superposition of Fock states,
\begin{equation}\label{eqPQ1}
\Dket{\psi} = \alpha_1\Dket{\phi_1} + \alpha_2\Dket{\phi_2}\,,
\end{equation}
with constant  $\alpha_i$. Now, without ``looking'' at the detector screen at $x=D$, let us insert a small screen directly behind slit $i$ to ``see'' if a quantum has passed through it. If so, then
\begin{equation}\label{eqPQ2}
N_{\op,i}\Dket{\psi} = \sum_{j=1,2}\alpha_jN_{\op,i}\Dket{\phi_j}=1\times\alpha_i\Dket{\phi_i}\,;
\end{equation}
and thus, consistent with the Born rule, the state $\Dket{\psi}$ has ``collapsed'' (or been ``reduced'') to $\Dket{\phi_i}$, with probability $\abs{\alpha_i}^2$. Indeed, replacement of $N_{\op,i}$ with one of its eigenvalues is tantamount to having performed a measurement. On the other hand, if we remove this slit detector and ``look'' only at the screen at $x=D$, we have no knowledge of which slit the source quantum may have passed through. The relevant number operator, therefore, must now be the aggregated number operator considering all possible quanta occupations, \viz,
\begin{equation*}
N_{\op}=N_{\op,1}+N_{\op,2}\,.
\end{equation*}
Then
\begin{equation}\label{eqPQ3}
\begin{split}
N_{\op}\Dket{\psi} &= \alpha_1N_{\op}\Dket{\phi_1} + \alpha_2N_{\op}\Dket{\phi_2}\\
&= \alpha_1N_{\op,1}\Dket{\phi_1} + \alpha_2N_{\op,2}\Dket{\phi_2}\\
& = 1\times\Dket{\psi}\,;
\end{split}
\end{equation}
that is, $\Dket{\psi}$, the superposition of states representing all possibilities at $x=0$, is an eigenstate of $N_{\op}$ belonging to eigenvalue $1$. Therefore, any measurement for $x>0$ in the presence of both slits is consistent with the interpretation of the superposition as describing a \textsl{single} quantum that, for a given source state, has been prepared by the slit screen for subsequent measurement, its result depending on where (and when) it is activated. To calculate the intensity at the detector screen appropriate to many repetitions of the measurement for equivalent source quanta we need to project the Fock space onto a Hilbert space of wave functions; the relevant wave function at any $(x,t)$ being $\psi(z,x,t)=\Dbraket{z,x,t}{\psi}$, with $\phi_i(z,x,t)=\Dbraket{z,x,t}{\phi_i}$ for $i=1,2$.\footnote{The shapes of the $\phi_i(z,x,t)$ must be specified, of course.} And then, again with appeal to the Born rule, the intensity measured along $z$ at $x=D$ is given by

\begin{equation}\label{eqPQ4}
\begin{split}
\abs{\psi(z,D,t)}^2 &= \abs{\alpha_1}^2\abs{\phi_1(z,D,t)}^2 + \abs{\alpha_2}^2\abs{\phi_2(z,D,t)}^2\\
&+  2\Re\bigl[{\alpha_1^*\alpha_2\phi_1(z,D,t)\phi_2^*(z,D,t)}\bigr]\,,
\end{split}
\end{equation}
displaying the well--known ``wave--like'' interference effect---due to the last term---of coherent superpositions.

The Fock space picture can also be applied more generally to superpositions of states, in particular those defining single--particle wave packets. Thus, for any number of orthogonal single--quantum basis states, we may construct an orthogonal Fock space basis
\begin{equation}\label{eqPQ10}
\Dket{\phi_i} = \Dket{0,\ldots,n_i=1,\ldots}
\end{equation}
for all $i$, where $n_i$ is the occupation number associated with each basis state, so that, as above, $N_{\op,j}\Dket{\phi_i}=1\times\Dket{\phi_i}\delta_{j,i}$. The Fock state corresponding to any linear superposition $\Dket{\psi}$ thus is
\begin{equation}\label{eqPQ11}
\Dket{\psi} = \sum_i w_i\Dket{\phi_i}\,,
\end{equation}
for any set of amplitudes $w_i$.\footnote{Fock states can be taken to be independent of time, while the amplitudes $w_i$ can absorb the time--dependence of the superposition; thus we can ignore the time here.} As in the two--slit analysis, it is easy to see that $\Dket{\psi}$ is an eigenstate of the aggregated number operator $N_\op = \sum_i N_{\op,i}$ belonging to eigenvalue $N=1$; \ie,
\begin{equation}\label{eqPQ12}
\begin{split}
N_\op\Dket{\psi} &= \sum_i w_i\sum_j N_{\op,j}\Dket{\phi_i}=\sum_{i,j}w_i 1\times\delta_{i,j}\Dket{\phi_i}\\  &=1\times\Dket{\psi}\,.
\end{split}
\end{equation}
This property is independent of the number of states in the superposition, and therefore, for well--behaved summations, can be generalized to continuous superpositions in the usual sense of defining an integral as the limiting value of a countably--infinite summation. Therefore, a wave packet expressed as, say, a linear superposition of a large number of plane waves is a quantum state of a single--particle in the relevant Fock space. The same holds true, of course, for any coherent linear superposition of fermionic states and, obviously, is consistent with the conventional scheme of SQM that a multi--particle wave function expressed as $\psi(x_1,x_2, \ldots, x_N)$ is an $N$--particle \textsl{pure} state.

Yet it might be suggested that the Fock space picture of wave packets is little more than window dressing. Recall, however, that in SQM the energy eigenstates of the \schrdngr equation for a single particle---certainly a fermion---undergoing simple harmonic motion usually are expressed in terms of multi--quanta (bosonic) Fock states, the very model for elementary QFT. An assertion that every $\psi(x)$ is necessarily a single--quantum state, therefore, may seem open to contextual interpretation, not to say confusion. In fact, however, the bosonic picture of oscillator states relevant to QFT stems from Maxwell's equations rather than the \schrdngr equation, and the manner in which fermionic states of the SHO can be recast as bosonic states is due to the fact that the \textsl{stationary} states of the \schrdngr and  Maxwell's equations satisfy identical differential equations after physical constants are redefined. It can be noted in this regard that Messiah \cite{Mess} remarks in his textbook treatment of the SHO that ``The [bosonic creation, destruction and number] operators were introduced to facilitate the solution of the  [stationary state] eigenvalue problem. If [in the \schrdngr equation] $\mathcal{H}$ is the Hamiltonian of a one--dimensional particle, these operators have no immediate physical significance.''\footnote{Further, in \cite{Mess}, in Vol.\ II, \S 5, p. 969, in the context of an introduction to QFT, one finds ``We have seen ... how the field can be quantized without the use of [oscillator] normal coordinates...[but] [their use] usually simplifies ... the interpretation of the theory of the quantized field.'' Also see \cite{Teller1}, Ch. 4, for an advanced discussion of such matters.} fermionic wave packets of interest to us in scattering problems are solutions of the time--dependent \schrdngr equation, and thus are not stationary states. So the context here is clear; superpositions of single--particle pure states are indeed---in any meaningful physical sense---pure states of single--quanta. A brief discussion of Fock space for wave packets from the point of view of creation and annihilation operators is offered in Appendix \ref{WSQ}.

\section{Scattering of 1-D Wave Packets}\label{1DWP}

\begin{itemize}
	\item 
\end{itemize}
\subsection{Stationary state plane wave solutions}\label{1DWP1}

We now turn to the specific interest of this work: 1-D wave packet scattering\footnote{Our emphasis here eventually will settle on ``dynamical'' (low energy) reflection, but the formalism applies to all the scattering.} from time--independent piecewise continuous potential energy barriers, compactly supported on the $x$--interval $0\le x<L$.\footnote{In mathematical language, ``support'' refers to regions of space on which a function, say $f(x)$, is non--zero. Compact support means support on a region of finite extent.}  This has been treated by Norsen, Lande, and McKagen \cite{NLMcK} using a similar approach to the one presented here, in which the wave packet is expressed in terms of plane wave solutions to the barrier problem, but relying mostly on numerical depictions of the scattered wave packet.  Jaimes-N\'{a}jera \cite{J-N} has addressed a similar problem but with focus on the time--dependence of the wave packet in the region of the barrier. Dimeo \cite{Dimeo1} has employed advanced methods of direct numerical solution of the \schrdngr differential equation for wave packets incident upon 1-D barriers, including time--varying barriers \cite{Dimeo2}. An early work by Ohmura \cite{Ohmura} presents a comprehensive description of wave packet scattering in the context of standard scattering theory, with emphasis on time--dependent effects. Muga \cite{Muga} also has presented an advanced analysis of wave packet scattering using M{\o}ller formalism, also focusing on temporal variations of the state in the interaction region. Our approach actually is quite straightforward and cannot claim originality in its basics. But our focus is to address the problem, often passed over, of generating the long--time limit amplitude for barrier reflection (and transmission) that directly pertain to the interpretation of typical scattering data. Because, as is well known, (most) wave packets spread continuously over time the (say) reflected wave packet no longer is localized in any meaningful sense by ``the time'' a reflected neutron is detected (as a click). As we will see, ``observable'' pure state amplitudes in the long--time limit turn out to be given (for a known incident wave packet and a known barrier potential) by a seemingly simple formula, perhaps one that might even be guessed, but here rigorously obtained and conceptually far from trivial.

To begin, we need to solve the time--dependent \schrdngr equation for barrier potential $V(x)$,

\begin{subequations}\label{eq.0}
\begin{equation}\label{eq.0a}
-\frac{\hbar^2}{2m}\partial_x^2\psi(x,t)+V(x)\psi(x,t|k)=\mi\hbar\partial_t\psi(x,t)\,,
\end{equation}
which we put into the form
\begin{equation}\label{eq.0b}
\bigl[-\partial_x^2\psi(x,t)+q(x)\bigr]\psi(x,t|k)=\mi\partial_{\beta t}\psi(x,t)\,,
\end{equation}
\end{subequations}
where, in commonly used notation, $q(x)=4\pi\rho(x)$, and $\rho(x)=(2m/\hbar^2)V(x)$ is the neutron scattering--length--density defining the barrier potential $V(x)$. Also in \eqref{eq.0b}, $\beta = \hbar/(2m)$, which has dimensions of $L^{-2}/T$. Equation \eqref{eq.0b} thus has dimensions of $L^{-2}$. In the analysis to follow we set $\beta=1$; in the end, true time can be restored by appropriate rescaling of dimensions.\footnote{In \eqref{eq.0b} we have chosen to link $\beta$ to $t$ as if it were, dimensions aside, a temporal scale factor. In other contexts, such as in Appendix \ref{sA}, it may be more natural to link it to the square of wave vector $k$, since for the free particle $E/\hbar=\beta k^2$. And, obviously, for the free particle phase angle $\beta(k^2 t)$, either association works.} The barrier, compactly supported on $0\le x\le L$, can now be represented by a set of (integer) $J$ contiguous scattering length pieces

\begin{equation}\label{eq.1}
q(x)=\{q_{0,x_1}(x),q_{x_1,x_2}(x),\dots,q_{x_{J-1},x_J=L}(x)\}
\end{equation}
for $x_j>x_{j-1}$ and with $q_{x_{j-1},x_j}(x)$ compactly supported on the interval $x_{j-1}\le x<x_j$.\footnote{
The 1-D problems of interest in a laboratory sense actually are geometrical projections of 2-D and 3-D problems onto the $x$--axis, say, in which the barriers become infinitely wide, uniform ``slabs'' satisfying $q(x,y,z)=q(x)$ within their spatial support. The \schrdngr equation (say for the 2-D case) then has factorable solutions as

\[[\partial_x^2\partial_y^2+q(x)]\psi(x,t|k)\me^{\mi k_y y-k_y^2 t}=\mi\partial_{\beta t}\psi(x,t)\me^{\mi k_y y-k_y^2 \beta  t}\,,\]
which lead at once to \eqref{eq.0} for $\psi(x,t|k)$, where the ``k'' appearing there is the $x$--component of the wave vector $(k_x,k_y)=\abs{\vec{k}}(\sin\theta,\cos\theta)$. The variation of the magnitude of \emph{this} $k$ can then be thought of as due to variation of $\theta$, the glancing angle of incidence. This picture allows us to think of detection of the scattering from the 1-D barrier as detection of specular scattering from the 2-D slab in the usual way, while taking advantage of the mathematical abstraction of the 1-D problem.}

Our immediate goal is formulate wave packet solutions of \eqref{eq.0}; but as we shall see, these can be developed in terms of familiar stationary state, plane wave solutions of the barrier problem---$\psi(x,t)=\psi(x,t|k)$ for given incident $k$---which we now briefly review. The stationary state solutions are given by linear superpositions of simple plane waves in the form

\begin{equation}\label{eq.2}
\psi(x,t|k)=\sum_{j=0}^{J+1}\psi_{j-1,j}(x,t|k)\,.
\end{equation}
The notation $\psi(x,t|k)$ may be read as $\psi(x,t)$ \textsl{given} wave vector $k$. Here $\psi_{-1,0}(x,t|k)$ is the part of the solution supported on all $x<0$ to account for the incident and reflected waves in the presence of $q_{-1,0}(x)=0$. In similar fashion $\psi_{J,J+1}(x,t|k)$ is the part of the solution supported on $x\ge L$ to account for the transmitted wave in the presence of $q_{J,J+1}(x)=0$ or for the case of a film attached to an infinitely thick substrate with $q_{J,J+1}(x)=q_b$ ($b$ for ``backing''). For the obvious benefit of simplicity, we now treat all scattering length density components as constants over their respective supports; \ie, for all $j$, $q_{x_{j-1},x_j}(x)=q_{x_{j-1},x_j}=q_j$. This allows for the rectangular description of any reasonable shape of $q(x)$ in the limit of refinement over smaller and smaller intervals of constant $q_j$, as needed. 

We label the three regions of the $x$--axis defined by the presence of the incident and reflected waves, the barrier, and the transmitted wave as I for $x<0$, II for $0\le x\le L$, and III for $x>L$. Then the stationary state solutions of \eqref{eq.0} can be defined by

\begin{subequations}\label{eq.3}
\begin{equation}\label{eq.3a}
\psi_\mathrm{I}(x,t|k)=\psi_{-1,0}(x,t|k)=\bigl[\me^{\mi kx}+r_{\pw}\!(k)\me^{-\mi kx}\bigr]\me^{-\mi k^2 t}\,,
\end{equation}
\begin{equation}\label{eq.3b}
\psi_\mathrm{II}(x,t|k)=\sum_{j=1}^{J}\bigl[\psi_{j-1,j}(x,t|k)=a_j(k)\me^{\mi n_j(k)k x} + b_j(k)\me^{-\mi n_j(k)k x}\bigr]\me^{-\mi k^2 t}\,,
\end{equation}
and
\begin{equation}\label{eq.3c}
\psi_\mathrm{III}(x,t|k)=\psi_{J,J+1}(x,t|k)=t_\pw\!(k)\me^{\mi kx}\me^{-\mi k^2 t}\,,
\end{equation}
\end{subequations}
where $r_\pw\!(k)$ and $t_\pw\!(k)$ in \eqref{eq.3a} and \eqref{eq.3c} are the plane wave reflection and transmission amplitudes, respectively.\footnote{In \eqref{eq.3} recall that now $\beta t = t$.} In this convention it is important to note that $r_\pw\!(k)$ in \eqref{eq.3a} is a function of the \emph{incident} $k$ but multiplies the reflected wave $\me^{-\mi k x}$. The function $n_j(k)$ appearing in the exponents of \eqref{eq.3b} and \eqref{eq.3c} is the index of refraction associated with each $q_j$, \viz,

\begin{equation}\label{eq.4}
n_j(k) = \sqrt{1-q_j/k^2}=\sqrt{1-k_{cj}^2/k^2}\,,
\end{equation}
where $k_{cj}=\sqrt{q_j}$. Thus one can also write $n_j(k)k$ as $\kappa_j(k)=\sqrt{k^2-k_{cj}^2}$. Notice that $n_j(k)$ does not appear in any of the $\me^{-\mi k^2 t}$ of \eqref{eq.3}, which, therefore, could be ``factored out'' of the summation in \eqref{eq.1}, since for stationary states the energy is constant over all $x$. It proves helpful for the wave packet scattering problem, however, to keep the $t$--dependent exponentials in place.

Finally, the $2J$ coefficients $a_j(k)$ and $b_j(k)$ in \eqref{eq.3b} along with the two ``interesting'' coefficients $r_\pw\!(k)$ and $t_\pw\!(k)$ must be determined by the $2(J+1)$ continuity requirements on $\psi(x,t|k)$ and $\partial_{x}\psi(x,t|k)$ at each of the $J+1$ interfaces defined by the piecewise continuous barrier. These coefficients are not strictly speaking unique, however, since their phases depend on free choices such as the position of the entire film along $x$ (or the placement of the coordinate origin) and---most importantly---the phase given to the incident plane wave, typically taken as unity as in \eqref{eq.3}. Indeed, as we will soon see, the behavior of wave packets constructed from these plane wave solutions is sensitive to the initial placement of the incident wave packet in space--time. Before turning to the wave packet problem, however, we note that while \eqref{eq.3b} for the barrier part of the wave function, $\psi_\mathrm{II}(x,t|t)$, is exact for the piecewise continuous barrier, its introduction of the $2J$ parameters $a_j(k)$ and $b_j(k)$ leads to what turns out to be unnecessary computational effort for large $J$. But by using the transfer matrix method, described in detail in the Appendix \ref{sB}, these coefficients can, in effect, be ``predetermined'' in a manner that leads to the reduction of $2J$ coefficients to just three easily found functions, $A(k)$, $B(k)$, and $C(k)$, depending only on $q(x)$, which completely determine $r_\pw\!(k)$ and $t_\pw\!(k)$ by simple formulas.\footnote{Actually there is a fourth function, $D(k)$, in this class but it can be determined from the other three by way of a unimodular property, as shown in Appendix \ref{sB}.}

\subsection{Wave packet solutions}\label{1DWP12}

A wave packet---by definition---is a general solution to the time--dependent \schrdngr equation, such as in \eqref{eq.0}; in short, it is a wave function, $\Psi(x,t)$, representing a single--particle \emph{pure state} at all $(x,t)$, as shown in Sec.\ \ref{PQFS}. As such, any wave packet can be mathematically described by a linear superposition of pure states, as discussed in Sec.\ \ref{PQSu}. Thus, in its most common representation (in continuum problems), a ``free'' wave packet, satisfying \eqref{eq.0} for $q(x)\equiv0$, can be constructed from a superposition of simple plane wave states $\phi_\pw(x,t|k)$---\ie, time--dependent wave vector eigenstates---such that

\begin{equation}\label{eq.40}
\Psi(x,t)=\Int \pco{k}\phi_\pw(x,t|k)dk
\end{equation}
with
\begin{equation}\label{eq.41}
\phi_\pw(x,t|k)=\me^{\mi(kx-k^2t)}
\end{equation}
and where $|p_\coh(k)|^2=\Pco{k}$ is the probability density for ``finding'' the value $k$ \textsl{in the state} $\Psi(x,t)$. The subscript ``\coh'' denotes coherency.\footnote{As Merzbacher \cite{Merz} points out: ``Again we emphasize that [in connection to wave packets] it is not permissible to consider $\Pco{k}$ as a measure of the relative frequency of finding various values of $k$ in a large \textsl{assembly} of particles. Instead it is the attribute of a \emph{single particle}. Naturally, when a theory predicts probabilities, the experimental verification of these probabilistic predictions requires the use of a large number of \emph{identical} systems [labeled by a given $k$ and $\Delta k$].''(Emphasis, minor notational changes, and the ending phrase have been added.) It should be noted that Merzbacher's discussion of wave packets is mostly limited to the free particle, as is the usual case in textbooks, but his succinct characterization as quoted here also applies to the general case, as well.} Later, in this context, we will replace  ``\coh'' with $\wp$ for ``wave packet''. Since every 
$\phi(x,t|k)$ satisfies \eqref{eq.0} (when $q(x)\equiv0$), the free wave packet also satisfies it for all well--behaved $\pco{k}$. By far the most commonly used model for $\pco{k}$ is a Gaussian function defined by a mean $k$--value, $\kbar$, and and a standard deviation $\Dk$. We show a well known explicit formula for the free particle wave packet in Appendix A. Then $\pco{k}\rightarrow \pco{k|\kbar,\Dk}$, with appropriate probability normalization

\begin{equation}\label{eq.42}
\begin{split}
\int_{-\infty}^\infty\abs{\pco{k|\kbar,\Dk}}^2dk = 
\int_{-\infty}^\infty\Pco{k|\kbar,\Dk}\,dk = 1\,.
\end{split}
\end{equation}
We display explicit forms of $\pco{k|\kbar,\Dk}$ and $\Pco{k|\kbar,\Dk}$ below. A much noted property of \schrdngr wave packets is their spreading in time or, equivalently, with distance traveled; the more localized the wave packet at any instance of time, say, $t_0$, the faster it spreads for $t>t_0$. As we will find, the effect of spatial spreading on measurements depends on exactly \emph{what} is measured.

For problems of interest, \viz, non--zero $q(x)$, we may construct $\Psi(x,t)$ in the basis given by the exact stationary state solutions of \eqref{eq.0}, $\psi_\st(x,t|k)$.\footnote{The idea of expanding wave packets with exact stationary state solutions has been given in a number of textbooks and articles, although rarely worked out beyond simple problems. The case of Fresnel reflection of a wave packet from a one--step barrier is discussed in  Vol.\ I CH.\ 3 \S 3 of Messiah's textbook \cite{Mess}, although with some hand waving, and the problem of properly defining reflectivity/transmission spectra for wave packets, as done here in Sec.\ \ref{1DWP2}, is not fully addressed.} For easier reading of equations, we suppress the dependence on the wave packet parameters, $\kbar$ and $\Dk$, except when it is helpful to show them. Therefore we may write
\begin{equation}\label{eq.45}
\begin{split}
\Psi(x,t)=\int_0^\infty p_\coh(k)\psi_\st(x,t|k)dk\,,
\end{split}
\end{equation}
where
\begin{equation}\label{eq.51}
\psi_\st(x,t|k)=\psi_\mathrm{I}(x,t|k)+\psi_\mathrm{II}(x,t|k)+\psi_\mathrm{III}(x,t|k)\\,
\end{equation}
as defined above in \eqref{eq.3}, each of these functions being compactly supported on its labeled region. The integration over $k$ in \eqref{eq.45} is limited to positive values, since the plane wave solution presumes the positive direction for incidence and the negative direction for reflection. However, for positive $\kbar$ not too small, we can extend the integration to negative values, relying on the weight function $p_\coh(k\gvn)$ to restrict integration to the positive $k$--domain. This packet may subsequently be resolved in terms of what we call the ``functional'' wave packets for the problem, \viz:

\begin{equation}\label{eq.52}
\begin{split}
\Psi(x,t)&=\Psi_\inci(x,t) + \Psi_\rf(x,t)
+\Psi_\barr(x,t)+\Psi_\tr(x,t)
\,,
\end{split}
\end{equation}
with ``$\inci$'', ``$\rf$'', ``$\barr$'', and ``$\tr$'' labeling the incident, reflected, barrier, and transmitted wave packets, respectively. These, in turn, are defined by

\begin{subequations}\label{eq.54}
\begin{equation}\label{eq.54a}
\Psi_\inci(x,t)=\theta(-x)\Int p_\coh(k)\phi_\pw(x-x_0,t|k)dk
\end{equation}
for the incident wave packet, centered at $x=x_0<0$ at $t=0$; and 
\begin{equation}\label{eq.54b}
\Psi_\rf(x,)=\theta(-x)\Int p_\coh(k)r_0(k)\phi(x+x_0,t|-k)dk
\end{equation}
for the reflected wave packet, initially ``located''---and thus effectively unobservable---at $x=-x_0=\abs{x_0}$; and 
\begin{equation}\label{eq.54c}
\Psi_\barr(x,t)=\theta(x)\theta(L-x)\Int p_\coh(k)\psi_\mathrm{II}(x,t|k)dk
\end{equation}
for the stationary barrier state, and, finally,
\begin{equation}\label{eq.54d}
\Psi_\mathrm{tr}(x,t)=\theta(x)\Int p_\coh(k)t_0(k)\me^{\mi k(n_b(k)-1)x}\phi_(x-x_0,t|k)dk
\end{equation}
\end{subequations}
for the transmitted packet. The explicit $x$--displacements $(x\pm x_0)$ appearing in $\phi_\pw(x,t|k)$ in \eqref{eq.54b} and \eqref{eq.54d} follow from the phase changes in the reflected and transmitted amplitudes induced by the displacement phase change in the incident plane wave; \ie, if $1\times \me^{\mi kx}\rightarrow\me^{-\mi kx_0}\times\me^{\mi kx}=\me^{\mi k(x-x_0)}$, then $r_0(k)\times \me^{-\mi kx}\rightarrow \me^{-\mi kx_0}r_0(k)\times\me^{-\mi kx}=r_0(k)\me^{-\mi k(x+x_0)}$, and similarly $t_0(k)\times \me^{\mi kx}\rightarrow t_0(k)\me^{\mi k(x-x_0)}$. These phase induced displacements spatially separate the initial centers of the reflection and transmission wave packet amplitudes from the initial center of the incident wave packet amplitude.

It is well to keep in mind that the four terms in \eqref{eq.52} describe a single neutron wave packet, just as  the corresponding piecewise continuous terms of the plane wave solution represent a single neutron state. The essential difference is that while $\psi_\pw(x,t|k)$ is a stationary state, $\Psi(x,t)$ is not.

\subsection{What is measurable?}\label{1DWP2}
With the wave packet $\Psi(x,t)$, eq.\,\eqref{eq.52}, in hand, we may ask what is measurable in that state. Obviously, in a ``position--measurement'' $\abs{\Psi(x,t)}^2dx$ gives us the probability of ``finding'' a particle at $x$ (\ie, in the interval $\{x,x+dx\}$) at time $t$. In practice, the registration of a space--time  event (we will call it a ``click''\footnote{Our characterization of measurement events as detector clicks is borrowed from Bohr, Mottelson, and Ulfbeck \cite{BMU}, although we do not necessarily take their view that clicks are ``fortuitous,'' \ie, without cause. Their argument is that a ``click'' is unpredictable because it is ``without history [or precursor],'' whereas causality conventionally is linked to an underlying temporal process. On the other hand, multitudes of clicks, they show within their theory, are statistically correlated to observables via constraints imposed by the breaking of space--time symmetry by the measuring apparatus, which involves the nature of the incident beam (\ie, the state prepared by the ``source''), the geometry and placement of hardware, and the placement and resolution of detectors.}) on a suitably placed detector is numerically converted into an angle that in turn, given the assumed known $k$--value for an incident particle, is converted into a wave vector subsequently interpreted as a scattered---say reflected---wave vector connected to the sample of interest.\footnote{In more exacting terms, the click of the detector located at a fixed position, say $X$, clearly signals a position measurement---in the sense that at time $t$ an event was registered at position $x=X$. (We do not here attribute any other connotation to the somewhat problematic word ``measurement.'') But in general, since a Hermitian operator $A_\op$ is given by $A_\op=\sum_i a_i\Dketbra{\phi_i}{\phi_i}$ in the basis of its eigenstates (assuming discreteness for simplicity), it follows that for any (smooth) function $f$, $f(A_\op)=\sum_i f(a_i)\Dketbra{\phi_i}{\phi_i}$. Thus, in these terms, a measurement of $x$ yielding $X$ can properly be viewed as a measurement of $\theta$ yielding $\Theta=f(X)$ without having to modify the apparatus. See, \eg,\ \cite{Haag}, Sec.\ I.1; also see Note 16 below. However, as mentioned in the text, converting $\Theta$ to a wave vector--value requires additional information.} In the literal 1-D problem the only possible angles for the $\vec{k}$ of concern are $0$ for incidence and transmission $(k>0)$, and $\pi$ for reflection $(-k<0)$. As described above in Note 2, however, the 1-D abstraction properly relates to the laboratory setup for measuring specular scattering from a 2--D (or 3--D) transversally smooth slab in such manner that a 1--D  $k$--value actually gives the magnitude of the normal (to the slab) component of a wave vector incident at glancing angle $\theta$. Variation of (1--D) $k$ therefore effectively generates the variation of incident plane wave angles that define a reflection or transmission spectrum.

The question remains, however; how do we relate these space--time clicks to the ``measured'' scattered intensities ultimately determined by a large number of such clicks for a given incident state? A familiar answer is given below at \eqref{SM}. First, however, we may appeal to the Heisenberg--Born rule: a detected--state event at $(x,t)$ producing $\phi(x,t)$  can be viewed as the result of a \emph{transition} to $\phi(x,t)$ from the state $\psi(x,t)$ that has evolved from the incident state $\psi(x,t_0<t)$, the probability of this transition being determined by its amplitude (in Dirac notation) $\Dbraket{\phi(t)}{\Psi(t)}$.\footnote{In loose terms, such a state transition is the result in hand (or conjectured result) of a measurement, without having to specify a mechanism or process responsible for it.} Thus, given $\Psi(x,t\gvn)$, the probability amplitude for a detected reflection assigned to a \emph{plane wave} of wave vector $-k$ is---up to an irrelevant phase---$r(k,t)=\Dbraket{\phi_\pw(t|-k)}{\Psi(t\gvn)}$, where $\Dbraket{x}{\Psi(t\gvn)}$ comes from \eqref{eq.54} and $\Dbraket{\phi_\pw(t|-k)}{x}=\phi_\pw^{*}(x,t|-k)$ comes from \eqref{eq.41}. That is,

\begin{equation}\label{eq.61}
\begin{split}
r(k,t)&=\int_{-\infty}^\infty\me^{\mi k x}\Psi(x,t)\,dx\\
&=\me^{\mi k^2 t}\int_{0}^\infty d\kp p_\coh(\kp)\me^{-\mi(\kp)^2 t} \me^{-\mi\kp x_0}\int_{-\infty}^0\,dx\bigl[\me^{\mi(k+\kp)x}+r_\pw(\kp)\me^{-\mi(\kp-k)x}\bigr]\\
&+\me^{\mi k^2 t}\int_0^L\me^{\mi kx}\Psi_\mathrm{II}(x,t)\,dx\\
&+\me^{\mi k^2 t}\int_0^\infty d\kp t_\pw(\kp)p_\coh(\kp)\me^{-\mi(\kp)^2 t} \me^{-\mi\kp x_0}\int_L^\infty\,dx \me^{\mi(\kp+k)x}\,.
\end{split}
\end{equation}
With the identity
\begin{equation*}
\int_0^\infty dx\,\me^{-\mi kx} = \pi\delta(k)+\frac{1}{\mi k}
\end{equation*}
in regions I and III, and a bit of manipulation, \eqref{eq.61}, becomes 

\begin{equation}\label{eq.62}
\begin{split}
r(k,t)&=\pi\me^{-\mi k x_0}p_\coh(k)r_\pw(k)\\
&+\me^{\mi k^2 t}\int_{0}^\infty d\kp \me^{-\mi\kp x_0}p_\coh(\kp)\me^{-\mi(\kp)^2 t}\bigl[\frac{1}{\mi(\kp+k)} + \frac{1}{\mi(\kp-k)}r_\pw(\kp)\bigr]
\\
&+\me^{\mi k^2 t}\int_0^L\me^{\mi kx}\Psi_\mathrm{II}(x,t)\,dx\\
&+\me^{\mi k^2 t}\me^{\mi kL}\int_0^\infty d\kp t_\pw(\kp)p_\coh(\kp)\frac{\me^{-\mi(\kp)^2 t}}{\mi(\kp+k)} \me^{-\mi\kp (L-x_0)}\,.
\end{split}
\end{equation}
The first term in \eqref{eq.62} is the immediate result of the appearance---from the $x$--integration---of $\delta(\kp-k)$, which can be satisfied in the $\kp$ integration since $\kp$ and $k$ both lie in the positive domain. On the other hand, the $\delta(\kp+k)$ that appear in regions I and III cannot be satisfied, so their contributions to the $\kp$ integrations vanish at once. The $x$--integration over region II is not as as easily simplified, since it depends on the detail shape of the barrier and involves the various index of refraction functions $n_j(k)$ that characterize the particle--barrier interaction. We will deal region II shortly. In order to account for the pole at $\kp=k$ that appears in the second line of \eqref{eq.62},
we everywhere slide $\kp$ ``slightly'' into the lower half of the complex $\kp$--plane as $\kp\rightarrow\kp-\mi\epsilon$, with $\epsilon>0$ and taken as small as necessary; \ie, formally, $\epsilon\rightarrow0^{+}$. Then we can use the Dirac identity

\begin{equation}\label{eq.63}
\begin{split}
\int_0^\infty d\kp\,\frac{f(\kp)}{\kp-k-\mi\epsilon} &= 
\pi f(k)+\mathrm{PV}\int_0^\infty d\kp\,\frac{f(\kp)}{\kp-k}\,,
\end{split}
\end{equation}
for any $f(\kp)$ well--behaved at $\kp=k$, where $\mathrm{PV}$ signifies the principal value---\ie, the integration which ``skips over'' $k$.\footnote{In the immediate vicinity, $k-a<\kp<k+a$, of the pole at $k$, the principal value integration, for $a>\epsilon$, can be written as 
\begin{equation*}
f(k)\bigl[\int_{k-a}^{k-\epsilon}\frac{d\kp}{\kp-k} + \int_{k+\epsilon}^{k+a}\frac{d\kp}{\kp-k}\bigr]\\
=f(k)\bigl[\int_{-a}^{-\epsilon}\frac{du}{u}+\int_{\epsilon}^{a}\frac{du}{u}\bigr] =0\,,
\end{equation*} for sufficiently well--behaved $f(\kp)$ near $\kp=k$. The Dirac formula can also be envisioned as integration along the real $\kp$--axis, except for ``going around'' the pole at $k$ with a semicircular contour of radius $\epsilon$, which gives one--half the Cauchy value for the pole term. The remaining integral is often called the Cauchy principle value.}  It is important to keep in mind that $\epsilon$, while an infinitesimal, is never zero. In practice, for sufficiently smooth $f(\kp)$ in the neighborhood of the pole, $\epsilon$ may be taken to be small but finite, typically the case in numerical implementations.

Eq.\ \eqref{eq.62} can now be written as

\begin{equation}\label{eq.64}
\begin{split}
r(k,t)&=2\pi\me^{-\mi k x_0}p_\coh(k)r_\pw(k)\\
&+\me^{\mi k^2 t}\int_{0}^\infty d\kp \me^{-\mi(\kp)^2 t}\mathcal{F}(\kp|k,x_0)\me^{-2\kp\epsilon t},
\end{split}
\end{equation}
where in the first term, $\pi$ has been replaced by 2$\pi$ as a consequence of the pole term from the $\kp$ integration in region I.\footnote{Curiously, this $2\pi$ also would have appeared if the $\kp$ integration in region $\rgn{I}$ had been taken over the entire $\kp$--axis.} In the second term $\mathcal{F}(\kp|k,x_0)$ stands for the detailed structure of the $\kp$ integrands in each $x$--region. The $t$--dependence of these integrands is common to each region, including region II, because of wave-function continuity. The $\epsilon^2$ dependence from $\exp[-\mi(\kp-\mi\epsilon)^2t$ in the $\kp$ integral has been safely ignored---but the first--order $\epsilon$ dependence shown in the final exponential factor of the integrand is, of course, crucial.

We are now interested in $r(k,t)$ in the limit $t\rightarrow\infty$, a common---if not near--universal---assumption in the definitions of observable amplitudes in scattering theory. Indeed, in standard $S$--matrix theory, given an initial state $\Dket{\Psi_i}$ defined at some $t=t_0$ (usually taken as``$t_0=-\infty$'') and assumed to be only source--dependent, eventual interaction with the scatterer produces 
$
\Dket{\Psi(t)}=S(t,t_0)\Dket{\Psi_i}
$ 
at finite $t$ and ultimately a ``final'' state  $\Dket{\Psi_f}=\Dket{\Psi(t\rightarrow\infty)}$ far from the scatterer.\footnote{In optics the observed state would be said to be characterized by its far--field behavior.} The resulting scattering amplitude is then defined  by 
\begin{equation}\label{SM}
S_{fi}=\lim_{t\rightarrow\infty}\Dbraket{\Psi_f}{\Psi(t)}=\Dbra{\Psi_f}S(\infty,t_0)\Dket{\Psi_i}\,.
\end{equation}
For our purposes, $\lim_{t\rightarrow\infty}r(k,t)$ represents $S_{fi}$ for reflection of the incident wave packet. The question then is how to treat the seemingly contradictory limits $\lim_{\epsilon\rightarrow0^+}\lim_{t\rightarrow\infty}$ in the exponent $2\kp\epsilon t$. Our answer is that, as already emphasized, $0^+$ is infinitesimal but non--zero ---$0^+=0+0^+>0$---moreover an infinitesimal that, in most practical cases, can be considered ``small'' but finite, as mentioned above; while, in keeping with formal scattering theory, $t$ can be as large as ``need be''---including $t=\infty$.\footnote{A simple way of thinking along such lines is that since $\epsilon$ and $t$ are completely independent variables, we can, for any $t$, approximate the principle value integrations to any desired degree of accuracy with suitably small $\epsilon$---and \emph{then} take $t\rightarrow\infty$, causing the integrals to vanish. Then we can return to the principle value integrations, making them even more accurate with still smaller $\epsilon$---and then, \emph{again}, take $t\rightarrow\infty$, causing the integrals to vanish, and so on for any degree of accuracy.} Thus, the second term in \eqref{eq.64} disappears in the limit of infinite $t$, and the observable reflection amplitude is given by the first term alone;\footnote{One can, of course, also appeal to the tendency of the ``violent'' oscillations of the phase factors $\me^{-2\kp{^2} t}$ as $t\rightarrow\infty$ to render the $\kp$--integrations in \eqref{eq.64} as vanishing, regardless of the value of $\epsilon$.} \viz,

\begin{equation}\label{eq.65}
\begin{split}
\lim_{t\rightarrow\infty}r(k,t)=2\pi\me^{-\mi k x_0}p_\coh(k)r_\pw(k)\,.
\end{split}
\end{equation}
And since what we ``measure'' at each $k$ is a reflected intensity $R_\coh(k\gvn)$, phase factors disappear, giving

\begin{equation}\label{eq.66}
R_\coh(k\gvn)=4\pi^2\gamma\abs{r(k,\infty\gvn)}^2 = 4\pi^2\gamma P_\coh(k\gvn)R_\pw(k)\,,
\end{equation}
where 
\begin{equation}\label{eq.68}
R_\pw(k) = \abs{r_\pw(k)}^2\,,
\end{equation}
$P_\coh(k\gvn)$ is defined in \eqref{eq.42}, and and where we have introduced an adjustable parameter $\gamma$ to facilitate computational comparisons, as in Figs.\ 5 and 6. It must be noted that the projection leading to \eqref{eq.65} presents the nuisance problem of producing a Dirac $\delta$--function in the limit that the wave packet ``becomes'' a plane wave $\Dket{k}$ with $\kbar\rightarrow k$; \ie, in the limit $\Dk\rightarrow 0$. Clearly we must honor the physical requirement that $R_\coh(k\gvn)\le 1$, which will be violated by most wave packet models, such as the Gaussian form, for sufficiently small values of $\Dk$. Thus we step around the problem by introducing $\gamma$; indeed, for pictorial convenience in the figures, which employ the Gaussian model, we choose 
\begin{equation}\label{eq.69a}
\gamma=\frac{\Dk}{4\pi^{\frac{3}{2}}}\,,
\end{equation}
to enforce $4\pi^2\gamma P_\coh(k\gvn)\le 1.$

One notices that the result for $R_\coh(k\gvn)$ in \eqref{eq.66} is also likely to be expected for an incoherent ``beam'' of plane waves---\ie, a mixed state---randomly distributed in $k$ values with probabilities equal to $P_\coh(k\gvn)$. Indeed, arguments have been made, for example by Ballentine in \cite{Ballentine}, ch.\ 9, and---in similar fashion, borrowing from Ballentine---Rauch and Werner in \cite{RandW}, ch.\ 12, that the scattering from a source producing even a single, reproducible wave packet is indistinguishable from that observed for the commonly assumed incoherent beam of plane waves.\footnote{Indeed, \cite{Ballentine}, p. 241, speaks of ``pointless speculation about the form of any \textsl{supposed} [emphasis added] wave function of an individual [particle]''; and ``the habit of associating a wave function with an individual [particle] instead of an ensemble''; and (in regard to a difference of opinion between \cite{Kaiser} and \cite{Cosma}) of ``the supposed wave packets of individual neutrons.'' The authors of \cite{RandW}, p.\ 393, put it a bit more gently, concluding that ``all observable quantities...are insensitive to the [difference between an incident wave packet and a beam of plane waves.]''} In Appendix D we discuss the force of Ballentine's argument.

Several figures show numerical examples for the double rectangular barrier in Fig.\! 1, with the plane wave reflection and transmission, $R_\pw(k)$ and $T_\pw(k)$ shown in Fig.\! 1, emphasizing the small--$k$ dynamical region.\footnote{It is common in scattering applications to distinguish between two regions of a scattering spectrum: the ``kinematic'' or ``Born region'' that is well described by first--order perturbation theory; and the ``dynamical'' region that requires higher orders and---at some point---the exact solution. In reflectometry, the dynamical region pertains to reflectivities rapidly approaching unity.} Here we have set all physical constants to unity, so scales for $k$ and $t$ are dimensionless. The $k$--distribution of the Gaussian model wave packet is, in this context, defined by

\begin{equation}\label{eq.69b}
(2\pi)^2\gamma\Bigl[\frac{1}{\sqrt{\pi}\Dk}\me^{-(k-\kbar)^2/(\Dk)^2}\Bigr]=\me^{-(k-\kbar)^2/(\Dk)^2}\,,
\end{equation}
where the bracketed function is normalized to unity and the result to the far right using the $\gamma$ in \eqref{eq.69a}. Fig.\! 3 displays the state function $|\psi(x,t)|$ representing the wave packet packet at $t=10$ (in black) and its incident shape (in gray) at $t=0$ for $\kbar=1.0$ and $\Dk=.4$. The deep ``wiggles'' can be attributed to interference of the of the overlapping incident and reflected portions of the wave packet, as discussed in \cite{NLMcK} and \cite{Dimeo1} with different barrier examples. Figs.\! 4a and 4b show the same  $|\psi(x,t)|$ having advanced to $t=150$. Fig.\ 4b, showing a close-up of the barrier region, illustrates the ``trapping'' effect of the ``hole'' in chosen barrier, even after a substantial amount of time; but since the hole depth is finite, eventually the weakly trapped wave packet segment ``leaks out.''  The relative smoothness of the reflected shape in Fig.\! 4a hides rapid oscillations of Re$\psi(x,t)$ and Im$\psi(x,t)$, as one expects for essentially trigonometric functions. 

Figs.\! 5 and 6 present $R_\coh(k\gvn)$ as defined by \eqref{eq.66} and \eqref{eq.69a} for several $\kbar$ values, with $\Dk$ equal to $0.025$ and $0.25$, respectively. These curves illustrate the ``non--specular'' scattering caused by the coherent distribution of $k$ values in the incident wave packet, with the specular components being associated with each $\kbar$.\footnote{We can refer to a reflection at $\kbar$ as ``specular'' in the sense that is corresponds to the normal use of the term in the $\Dk\rightarrow0$ limit.} One sees in Figs.\! 5 that small $\Dk$ gives nearly specular reflection, as indicated by narrow spectral ``lines.'' In the limit $\Dk\rightarrow0$, incident wave packets centered on a given $\kbar$ behave, of course, as incident plane waves ``centered'' on the same $\kbar$, so then only specular reflection---$k_\textrm{out}=-k_\textrm{in}$ in 1D---is observed. Non--specular reflection increases with increasing $\Dk$, but we notice that its ``weight'' goes to zero wherever $R_\pw(k)\rightarrow0$. 

Fig.\ 7 shows $\abs{r(k,t)}$ for $t = 150,\ 30,\ \textrm{and}\ 10$ as obtained from numerical $x$--integration\footnote{The numerical $x$--integration was done with commercial mathematical software using Gauss--Lobatto--Kronrod quadratures.} of \eqref{eq.61}, showing quite good agreement at $t=150$ with the formula in \eqref{eq.65} for $|r(k,$``$t=\infty$'')$|$. (Vertical scales in the figure where chosen to ease comparison.) As the figure also reveals, agreement at small $k$ values is subject to small but noticeable oscillatory deviations. While partly due to numerical rounding, such behavior may also suggest that plane waves ``within the packet'' move more slowly as $k$ gets closer to zero, thus covering less distance in a given finite time interval. In effect, as $k$ becomes smaller, any large but finite $t$ behaves as if it were smaller as well, pushing further away the asymptotic limit. Perhaps most importantly, Fig.\ 7 emphasizes that wave packet scattering from the barrier is indeed inherently dependent on the observation time $t$, a dependence that effectively disappears only in the long--time limit.

$R_\coh(k\gvn)$, as defined by \eqref{eq.66}, assumes that every wave packet impinging on the barrier has the same values of the parameter pair, $\kbar$ and $\Dk$; in that sense, the incident beam could be likened to a monochromatic plane wave beam. In this work we do not delve into the difficult topic of wave packet formation and take it for granted that such ``monochromatic" wave packet beams are possible. However, one can also consider a wave packet beam characterized by uniform $\Dk$ but incoherently variable $\kbar$, by analogy to an incoherent beam of plane waves of different $k$ values. We can, indeed, define a class of wave packets in this manner, the class being identified by its common value of $\Dk$ while members of the class have different values of $\kbar$. In real space, all members of such a class thus share the same degree of localization---as measured, say, by ``width" or standard deviation, while individual members may propagate with different group velocities. Figs.\, 5 and 6 show reflectivities in the neighborhoods of several values of $\kbar$ for two such incident beam classes, corresponding respectively to $\Dk=0.025$ and $0.25$. The case of Figs.\, 5 indicates an effective approach to specular reflection in a class on the verge, one might say, of behaving plane--wave like. We note, however, that near the specular zeros, the off--specular reflections behave asymmetrically about the nominal $k=\kbar$.  

Source--independent instrumental resolution defines yet another class of wave vector values---distinct from a source--dependent wave packet class---we will call $\{\km,\Delkinst\}$, where $\km$ refers to the $k$--value we \textsl{assign} to the position of a detector click and $\Delkinst$ characterizes the spread of such assignments associated with imperfect detectors; \ie, the $\km$ we report could have been assigned to a different, presumably nearby, location of the detector click. Such associations generally are mathematically expressed by statistical resolution functions specifically linking reported $k_\meas$--values to a set of ``true'' $k$--values---\ie, $k$--values within a coherent population---say as $P_\inst(\km-k|\Delkinst)$---a near--``perfect'' detector corresponding to $\Delkinst\rightarrow0$, and thus leading to $P_\inst(\km-k|\Delkinst)\rightarrow\delta(\km-k)$. Then, using \eqref{eq.66}---with ``$\coh$'' now notated as ``$\wap$''---a \textsl{measured} wave packet reflectivity spectrum for a source defined by a given incident wave packet class and a $\kbar$ within that class can be expressed as

\begin{subequations}\label{eq75a}
\begin{equation}\label{eq.75aa}
\begin{split}
R_{\meas/\inst}(\km|\Delkinst,\kbar,\Delkwp) &= \int_0^\infty\,dk P_\inst(\km-k|\Delkinst)R_{\Wp}(k|\kbar,\Delkwp)\\
&=\int_0^\infty\,dk P_{\meas/\inst}(\km,k|\kbar,\Delkinst,\Delkwp)R_\pw(k)
\end{split}
\end{equation}
with
\begin{equation}\label{eq.7ab}
P_{\meas/\inst}(\km,k|\kbar,\Delkinst,\Delkwp) = 4\pi^2\gamma P_{\inst}(\km-k|\Delkinst)P_{\Wp}(k|\kbar,\Delkwp)
\end{equation}
\end{subequations}
accounting, independently, for both the wave packet class and instrumental resolution.

Appropriately similar results are obtained, of course, for the case of transmission measurements, where we begin with the amplitude defined by $t(k,t)=\Dbraket{+k}{\Psi(t)}$. It is well to point out here that reflectivities measured as  $R_{\meas/\inst}(\km|\Delkinst,\kbar,\Delkwp)$ will tend to mask, \ie ``smear'', the distinction between specular and non--specular wave packet reflectivities exhibited in Figs.\! 5 and 6 when $\Delkinst$ becomes comparable to $\Delkwp$. Thus, in order to clearly observe wave packet effects in spectra from wave packets in classes of small $\Delkwp$, which---depending on the details of the barrier---may behave similarly to a plane wave sources, suitably sharp instrumental resolution is likely to be required. 

For a simple illustration of instrumental resolution in \eqref{eq.7ab} we use the (un--normalized) Gaussian function\footnote{Thus here, $\Delta k=\sqrt{2}\sigma$ for each manner of wave vector spread, where $\sigma$ represents the statistical definition of standard deviation.}
\begin{equation}\label{eq.7ac}
P_{\inst}(\epsilon|\Delkinst) = \me^{-\epsilon^2/\Delkinst^2}\,.
\end{equation}
Fig.\! 8 shows an example of $R_{\meas/\inst}(\km|\Delkinst,\kbar,\Delkwp)$ for the wave packet class illustrated in Fig.\! 5 but now with $\Delkinst=0.05$ (not zero). The vertical scale is chosen for illustration purposes. Thin lines show the individually broadened lines associated with each ``measured'' $k$--value, while heavy lines show the result of overlapping. The inset for the $\kbar=0.75$ member of the given wave packet class uses a heavy lines for the instrumentally broadened line a heavy dashed line and the ``perfect'' wave packet line broadened by non--specular scattering in Fig.\! 5. One notices that the zeros at $\kbar=0.5$ and $1.0$---otherwise preserved by the wave packet, as in Fig.\! 5 and 6---are now gone, and observable line broadening is significantly enhanced by the assumed, somewhat loose, instrumental resolution.
 
To conclude this section, we return to \eqref{eq.65}, which might appear to ``ignore'' the fact that \schrdngr wave packets, $\Psi(x,t)$ spread unrelentingly in time. Yet \eqref{eq.65} and \eqref{eq.66} express $r(k,t\rightarrow\infty)$ in terms of the wave packets generated by $p_\coh(k|\kbar,\Delkwp)$ independently of time, and which, therefore, can be associated with the $k$ distribution in the \textsl{incident} wave packet. However, the measurements of \emph{reflectivity} (or transmission) spectra, as generally interpreted, are not measurements of $\Psi(x,t)$; rather, they are treated as measures of detected intensities as defined by the moduli of its projections onto wave vector eigenstates corresponding to reflected (or transmitted) plane waves.\footnote{In SQM more generally, given two (normalized) pure states $\Dket{\psi_1}$ and $\Dket{\psi_2}$, the probability of transition from $\Dket{\psi_1}$ to $\Dket{\psi_2}$ is defined by $\abs{\Dbraket{\psi_2}{\psi_1}}^2$. The connection of such a transition to a measurement involves additional mathematical and interpretive concerns.} As emphasized by the complexity of its derivation, the apparent simplicity of the result for the reflection amplitudes produced by $\Psi(x,t)$ does indeed require the limit $t\rightarrow\infty$, because it is only in this limit that the ``past history'' of the wave packet plays no significant role in the measurements.\footnote{Also see Appendix D, Eqs. \eqref{eq.WPTA10} and \eqref{eq.WPTA11}.}

\begin{figure}
\begin{minipage}[c]{0.4\linewidth}
\includegraphics[width=\linewidth]{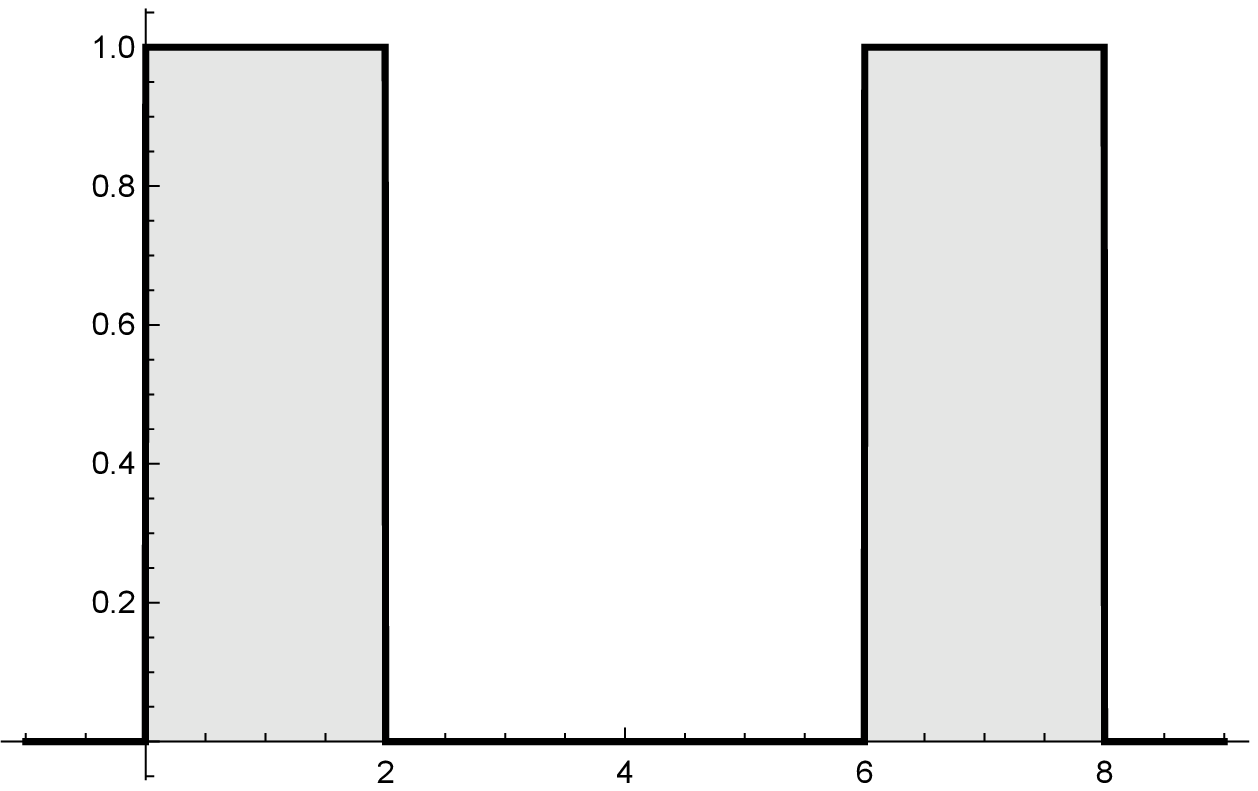}
\caption{Barrier for subsequent figures.}
\label{fig:arXfig1}
\end{minipage}
\hfill
\begin{minipage}[c]{0.4\linewidth}
\includegraphics[width=\linewidth]{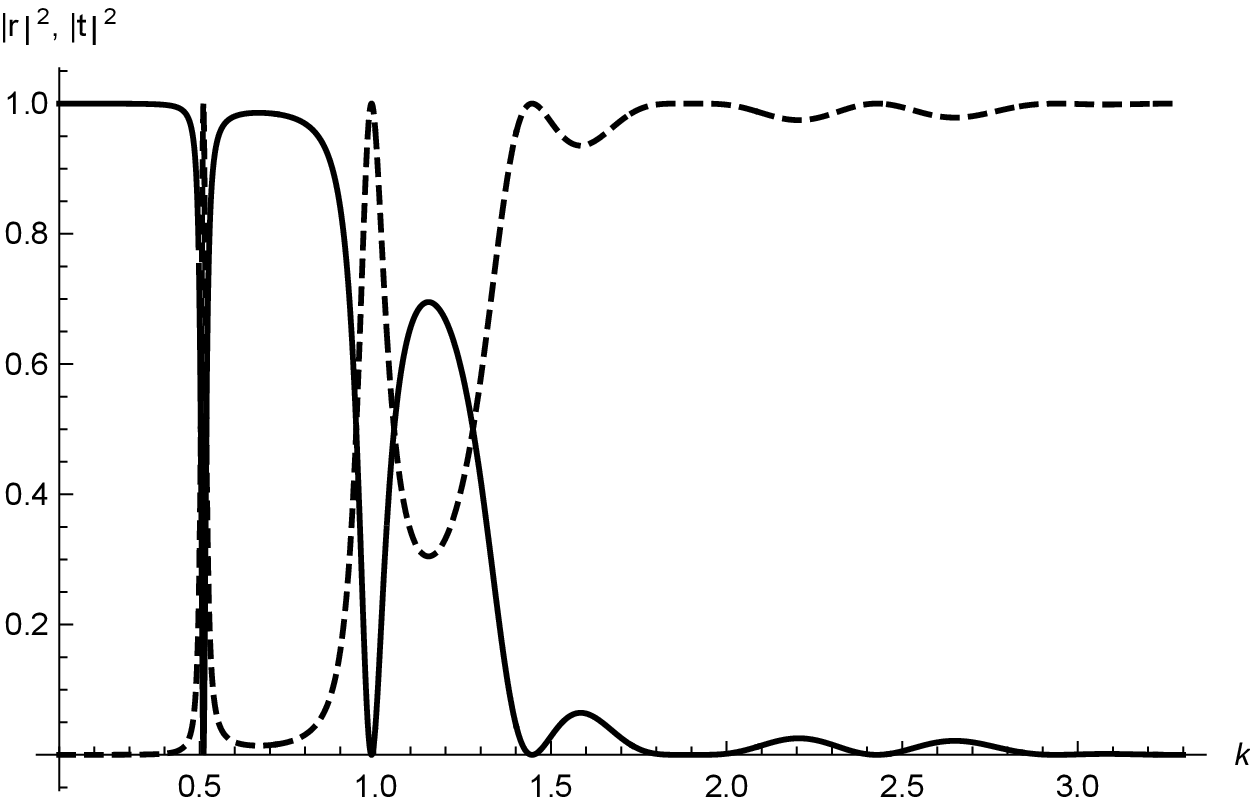}
\caption{Plane wave $|r|^2$ (solid) and $|t|^2$ (dashed) for barrier in Fig.\ \ref{fig:arXfig1}\hspace*{\fill} }
\label{ arXfig2}
\end{minipage}%
\end{figure}

\begin{figure}
\begin{minipage}[c]{0.4\linewidth}
\includegraphics[width=\linewidth]{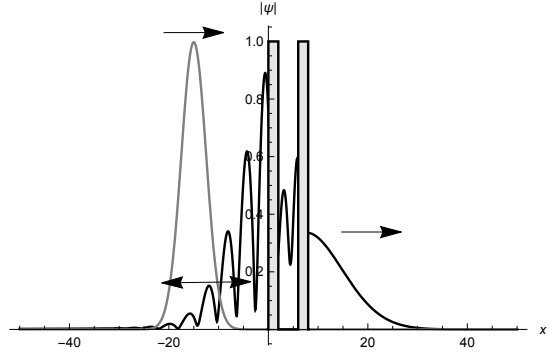}
\caption{Wave packet (black) near barrier at $t=10$.\hspace*{\fill} $\kbar=1.0$, $\Dk=.4$. Incident wave packet (light black) shown at $t=0$.\hspace*{\fill}}
\label{fig:arXfig3}
\end{minipage}
\hfill
\end{figure}

\begin{figure}
\begin{minipage}[c]{0.4\linewidth}
\includegraphics[width=\linewidth]{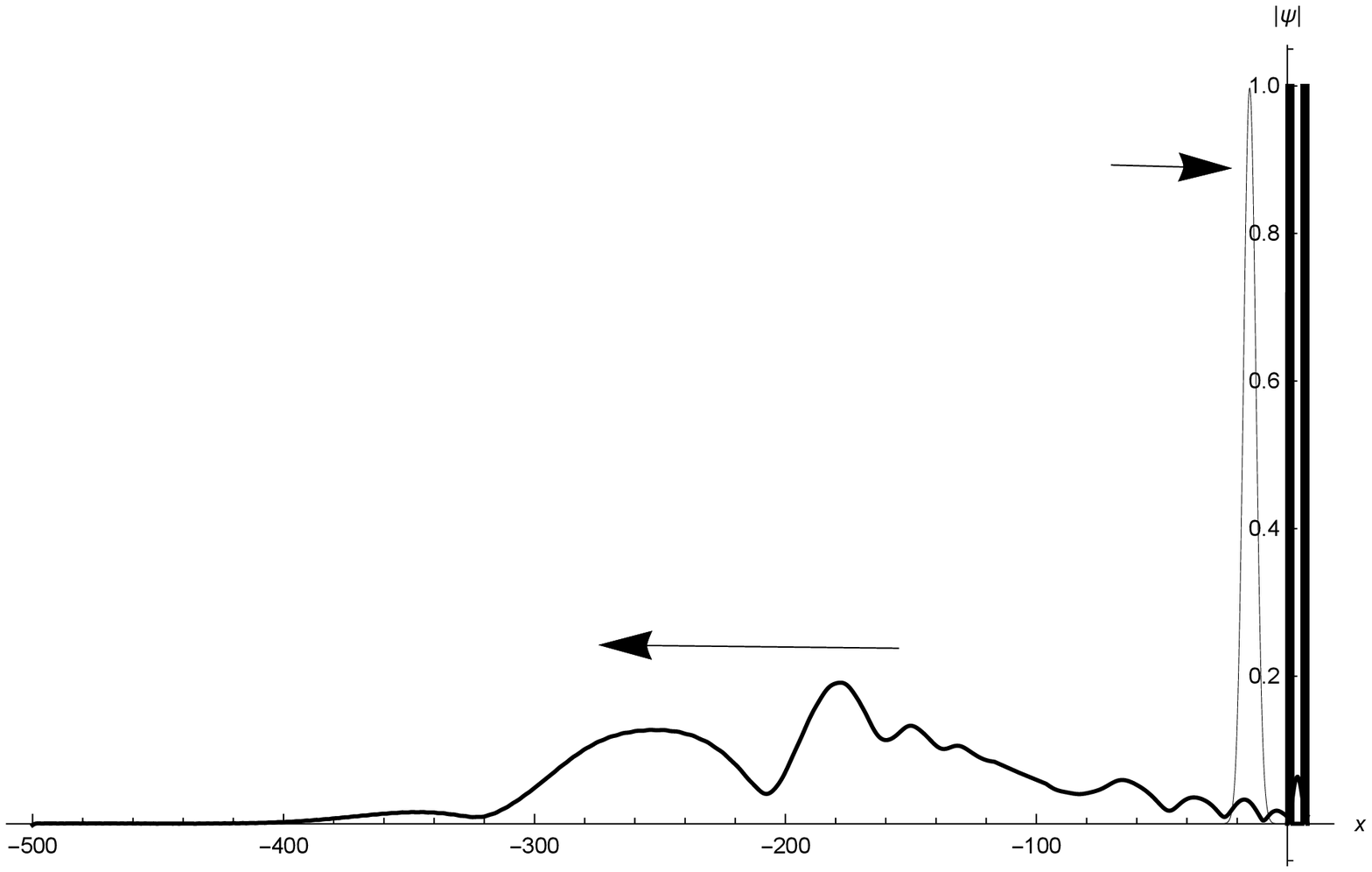}
\caption{Wave packet (heavy black) reflected from barrier (at right) at $t=150$. Incident wave packet (light black), for $\kbar=1.0$, and $\Dk=.4$, is shown near barrier at $t=0$\hspace*{\fill}.}
\label{fig:arXfig4a}
\end{minipage}
\hfill
\begin{minipage}[h]{0.4\linewidth}
\includegraphics[width=\linewidth]{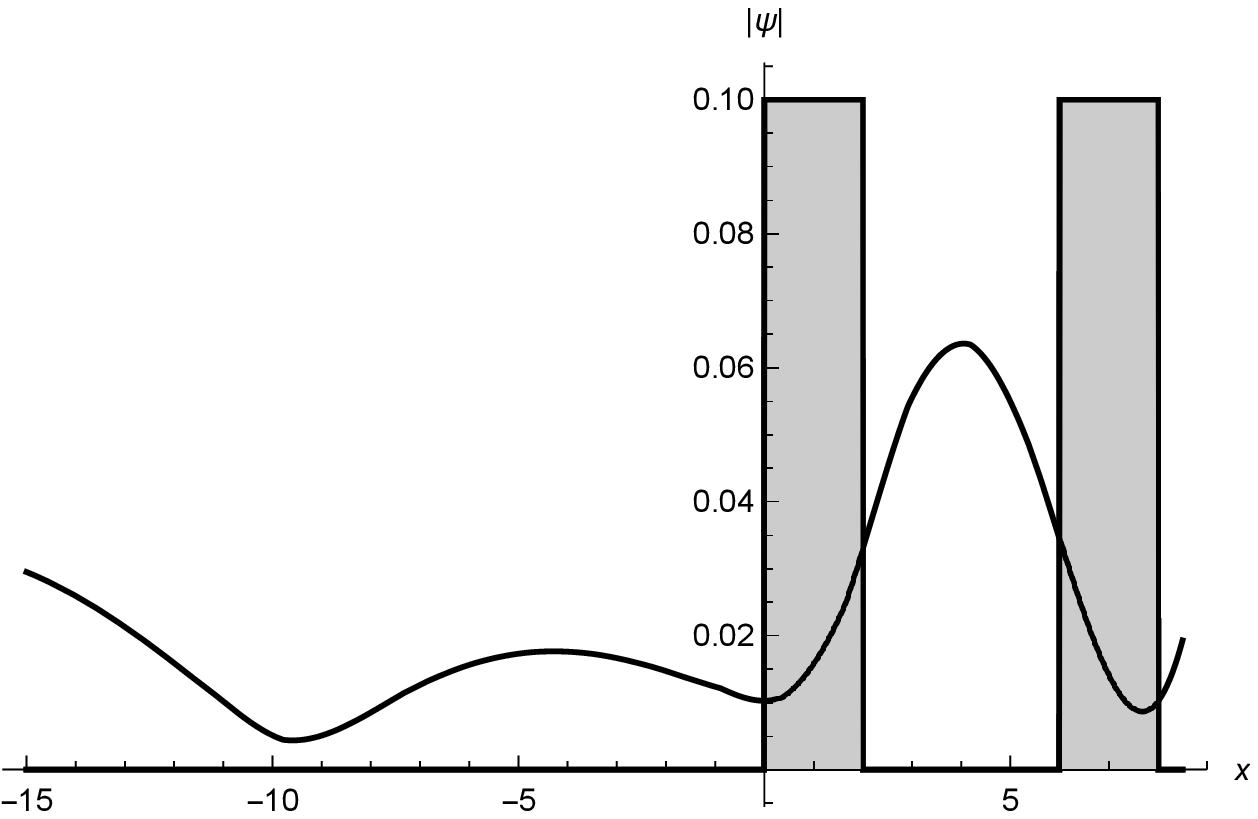}
\caption{Closeup of Fig.\ 4.\hspace*{\fill}.}
\label{fig:fig4b}
\end{minipage}
\end{figure}%

\begin{figure}
\begin{minipage}[t]{0.4\linewidth}
\includegraphics[width=\linewidth]{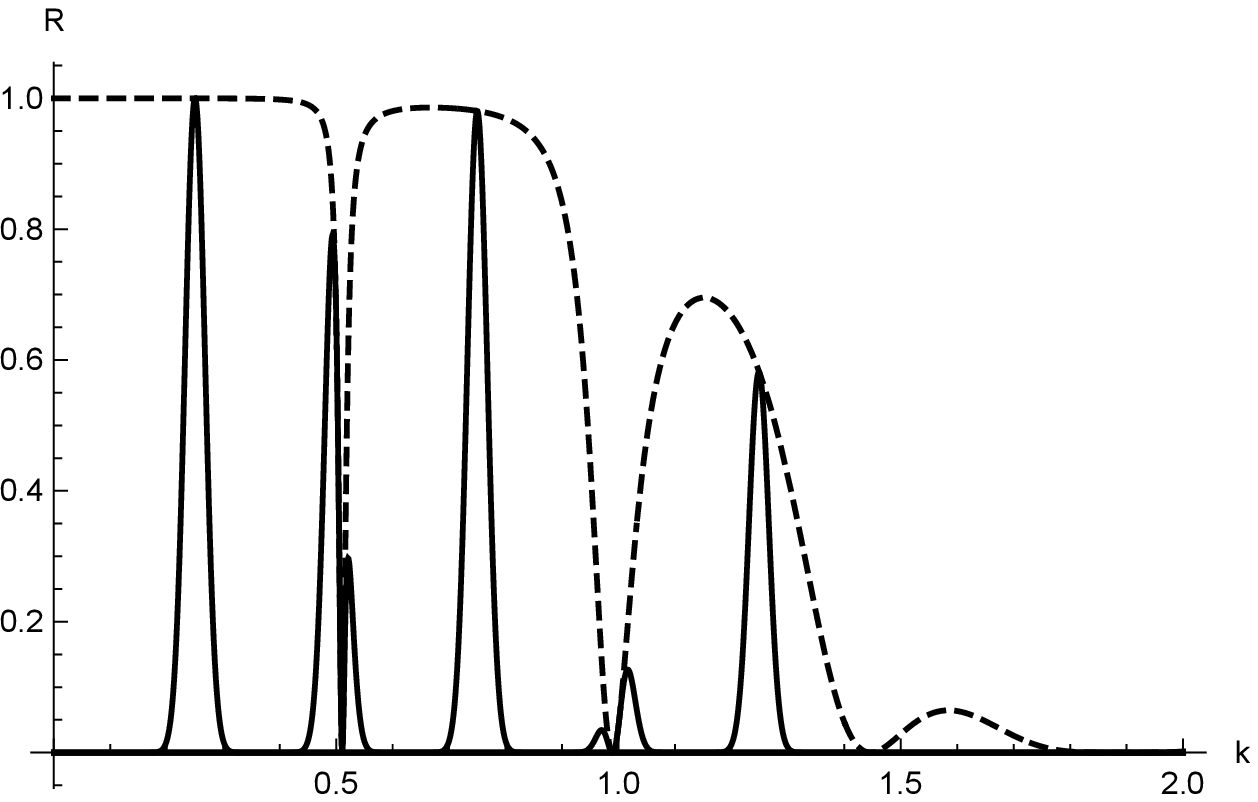}
\caption{Reflectivities from Fig.\ 1 barrier for wave packets with $\Dk=0.025$ and values of $\kbar = 0.25, 0.5, 0.75, 1.0, 1.25$. $R_{\pw}(k)$ is dashed.}
\label{fig:arXfig5}
\end{minipage}
\hfill
\begin{minipage}[t]{0.4\linewidth}
\includegraphics[width=\linewidth]{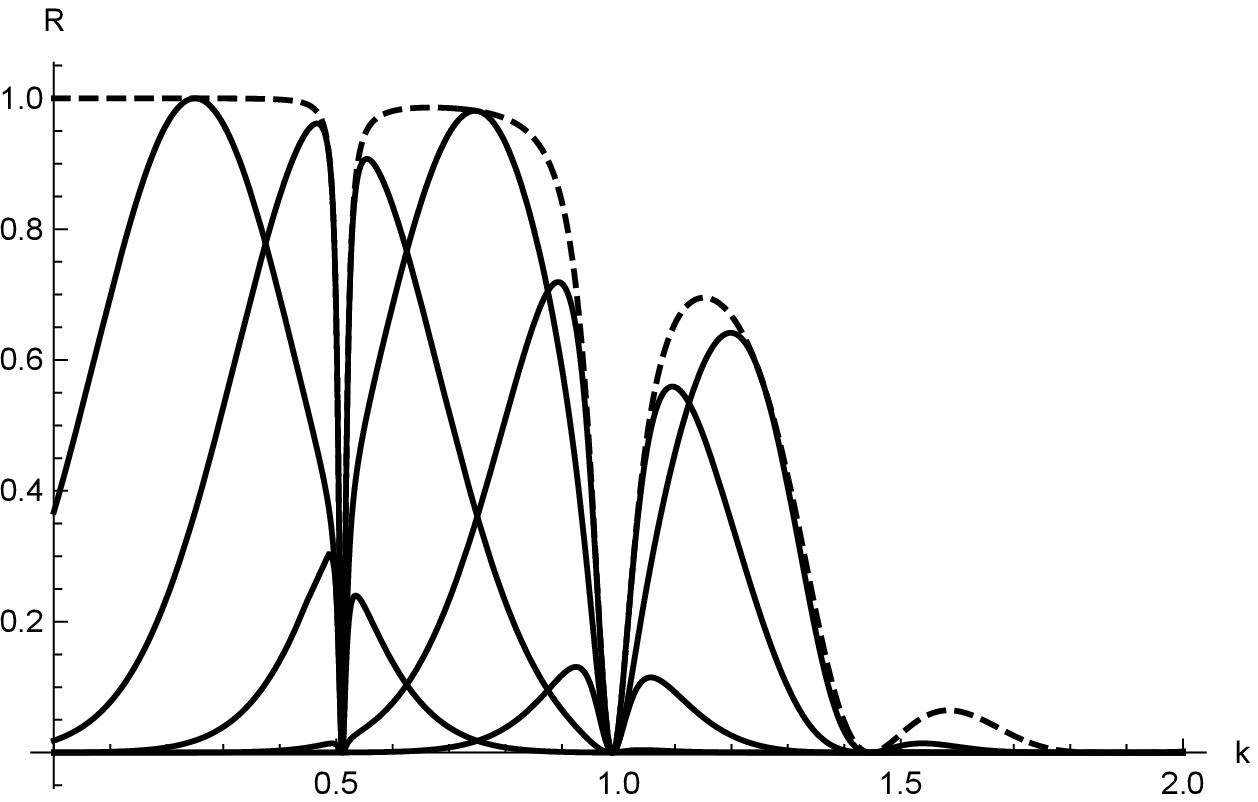}
\caption{Reflectivity from Fig.\ 1 barrier for\hspace*{\fill} wave packets defined by $\Dk=0.25$ and the $\kbar$ values in Fig.\,5.\hspace*{\fill}}.
\label{fig:arXfig6}
\end{minipage}
\end{figure}

\begin{figure}
\begin{minipage}[t]{0.4\linewidth}
\includegraphics[width=\linewidth]{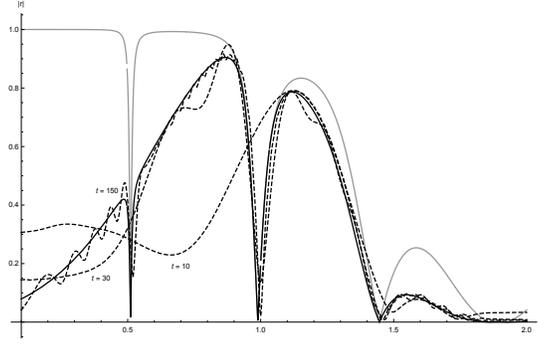}
\caption{\hspace*{-.2em}Plot of $|r|$ \textsl{vs} $k$ for $\Dk=0.4$, and $\kbar=1$. 
Solid line: $|r(k)|=|\Dbraket{-k}{\psi}|$ from formula (26). 
Dashed lines from calculating \hspace*{-.3em}\hspace*{-.3em}(24) 
for $t=150$, $t=30$, and $t=10$, with \hspace*{-.2em}$x_0=-15$, showing approach to $t\rightarrow\infty$. }
\label{fig:arXfig7}
\end{minipage}
\end{figure}

\begin{figure}
\begin{minipage}[t]{0.4\linewidth}
\includegraphics[width=\linewidth]{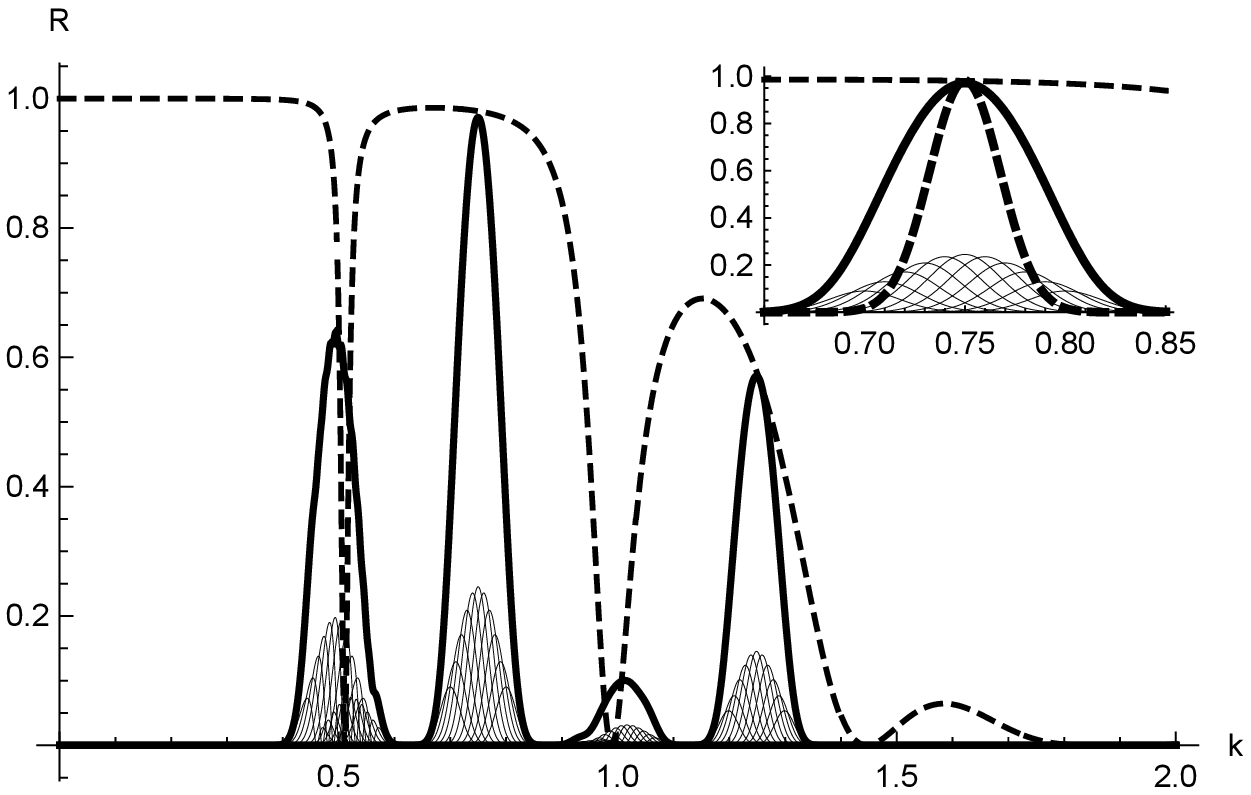}
\caption{Plot of $R_{\meas/\inst}$ for the same class, $\Dk=0.025$, and $\kbar$ set as in Fig.\! 5, but now with instrumental resolution $\Delkinst=0.05$. \hspace*{.8em}See text below eq.\,\eqref{eq75a}.\hspace*{\fill}}
\label{fig:arXfig8}
\end{minipage}
\end{figure}

\begin{acknowledgements}
The author is grateful to longtime collaborator C.\ F.\ (``Chuck'') Majkrzak for sharing his deep insights on neutron scattering and for his constant encouragement of the work reported here. 
\end{acknowledgements}

\begin{appendix}
\numberwithin{equation}{section}

\section{The Gaussian wave packet}\label{sA}
We begin with a bit of history. While \schrdngrs 1926 papers made many references to wave packets (at the time often called ``wave groups'') his emphasis was on stationary states associated with atomic orbitals. It appears that the first comprehensive mathematical treatment of free (\ie, unbound) particle wave packets satisfying the time--dependent \schrdngr equation came soon after (1927) in (what is today) a relatively unknown paper by physicist C.\ G.\ Darwin \cite{Darwin} (grandson of Charles) that is remarkably modern in its presentation.

For easy reference we display the free particle wave packet as defined in \eqref{eq.40} for the standard Gaussian model, specified here as

\begin{equation}\label{eq.sA.1}
p_\coh(k\gvn) = \me^{-\frac{(k-\kbar)^2}{2(\Dk)^2}}\,.
\end{equation}
For simplicity here, we ignore proper normalization in order to employ the familiar normalized Gaussian form of the exponential appropriate to the definition of $\Dk$ as a standard deviation. When proper normalization is required of the squared modulus $P_\coh(k\gvn)$, then $(\Dk)^2$ in \eqref{eq.sA.1} must be replaced by $2(\Dk)^2$, an easily overlooked nuisance. For additional contact with most textbook expositions, we allow the form of $p_\coh(k\gvn)$ for all values of $k$. As discussed above, however, a wave packet description of the incident particle in a scattering context must take into account that the wave vectors associated with incident particle states should be consistent with the direction defined by such incidence. Specifically, in the one dimensional case, if the direction of incidence is defined by positive wave vectors, negative wave vectors must not contribute to an incident wave packet. So then, the right hand side of \eqref{eq.sA.1} should be multiplied by the Heaviside theta function $\theta(k)$, which vanishes for $k<0$, in order to maintain a proper description of incidence. Such a restriction becomes vital when exact plane wave scattering states become the basis for the wave packet solution of the \schrdngr equation appropriate to the scattering setup, as discussed in the text. This also leads, however, to a rather more complicated formula for the wave packet. As a practical matter, we may loosen the incidence requirement by assuming, say, that $\kbar>3\Dk$, as nearly would be the case for  almost--plane--wave--like wave packets; the Gaussian form itself then (almost) enforces the restriction to $k>0$.

With a bit of additional notation, \eqref{eq.40} and \eqref{eq.41} therefore give (without normalization)

\begin{subequations}\label{eq.sA.3}
\begin{equation}\label{eq.sA.3a}
\Psi(x,t) = c(t)\me^{-\frac{(X-2\beta\kbar T)^2}{2\sigma(t)^2}}\me^{-\mi\chi(x,t)}\,,
\end{equation}
where
\begin{equation}\label{eq.sA.3b}
\sigma(t) = \sqrt{\sigma_0^2 + 4\beta^2T^2/\sigma_0^2}= \sigma_0\sqrt{1 + \frac{\Dk^{\prime 2}}{k^2}\frac{X^2(T)}{\sigma_0^2}}\,,
\end{equation}
\begin{equation}\label{eq.sA.3c}
\chi(x,t) = 4\Bigl(\kbar X-(k^2-X^2/\sigma_0^4)\beta T\Bigr)\,,
\end{equation}
and
\begin{equation}\label{eq.sA.3d}
c(t) = \frac{\Dk}{\sqrt{1+2\mi\beta\Dk^2T}}\,,
\end{equation}
\end{subequations}
with $\sigma_0 = (\Dk)^{-1}$, and $X=x-x_0$ and $T=t-t_0>0$. (Typically, in textbook discussions, $x_0=t_0 = 0$.) In addition we have restored the constant $\beta\ne 1$, that appears in \eqref{eq.1}; we note that $\beta t$ has dimension $\mathrm{L}^2$.\footnote{For neutrons, $\beta t(sec) \approx 3.15\time 10^4\mu m^2$. } Also, in \eqref{eq.sA.3b}, $X(t)=v_g T$ is the distance traveled by the wave packet in time $T$, where $v_g=2\beta k$ is the group velocity. Thus in \eqref{eq.sA.3a} and \eqref{eq.sA.3b} we see the ``famous'' spreading of the (massive) free wave packet with increasing $t$ (or X(T)), the rate of spreading increasing with smaller initial spatial localization (\ie, with greater $\Dk$).\footnote{Spreading with time is not specific to the Gaussian model. Messiah \cite{Mess1}, in fact, derives (with some hand waving) a formula very similar to \eqref{eq.sA.3b} without assuming a Gaussian.}

Non--spreading wave packets---called solitary waves or solitons---are characteristic of massless particles, a well--known mathematical result of the linear $k$--dependence of their kinetic energy. solitons are possible for massive particles, however, in the relativistic realm as special solutions, for example, of the 3--D Klein--Gordon equation; see \cite{Mackinnon} and \cite{Mosley1}. As regards solutions of the \schrdngr equation, however, we are aware of only one example of a true soliton---but only in 3--dimensions, as a spherically symmetric wave packet \cite{Mosley2}.

Be aware that the last expression in \eqref{eq.sA.3b} notwithstanding, $\sigma(t)$  is independent of $k=\kbar$, the mean, ``group,'' or ``drift'' wave vector of the wave packet, implying that a 2-- or 3--dimensional wave packet spreads in time in each of its geometrically defined dimensions, dependent only on the value of $\Delta \kp$ for each dimension. Specifically, for a 2--D $\{x,y\}$ wave packet with longitudinal (\ie, drift) dimension along $x$ and transverse dimension along $y$, and with the Gaussian model applying along each axis, we now have \eqref{eq.sA.3} along $x$, with $\sigma(t)\rightarrow\sigma_{x}(t)$, $k\rightarrow k_x$, etc.; while along $y$, $\sigma_0(t)\rightarrow\sigma_{0y}(t)=1/\Dk_y^\prime$, $k\rightarrow k_y=0$, etc., and now with
\[\sigma_y(t) = \sqrt{\sigma_{0y}^2 + 4\beta^2T^2/\sigma_{0y}^2}= \sigma_{0y}\sqrt{1 + \frac{\Dk_y^{\prime 2}}{k_x^2}\frac{X^2(T)}{\sigma_{0y}^2}}\,,
\]
the last formula expressing the broadening along $y$ in terms of the distance traveled along $x$. It is interesting to notice that since the rate of broadening along orthogonal dimensions increases with the degree of initial localizations along the axes, a cigar--shape (elongated) wave packet eventually evolves into a pancake--shaped (flattened) wave packet as it expands; similarly, an initial ``pancake'' expands into a ``cigar.''

Wave packet spreading is a purely mathematical result of the quadratic form of kinetic energy for particles of non--zero mass, \viz, $E/\hbar =\beta k^2$. It is sometimes rationalized by appeal to the uncertainty principle, which, however, could possibly suggest the time--dependent constraint $\sigma(t)\sim 1/\Dk(t)$. Such behavior, however, depends on the subtlety of whether we are considering the state $\Psi(x,t)$ or its modulus $\abs{\Psi(x,t)}$. For example, with \eqref{eq.42}, \eqref{eq.41} can be expressed as the Fourier transform (FT) of $p_\coh(k)\me^{-\mi k^2 t}$, which, for Gaussian $p_\coh(k)$, as in \eqref{eq.sA.1}, spreads according to \eqref{eq.sA.3}. Then, except for normalization, the distribution of component--$k$ values in $\Psi(x,t)$, as defined by its \emph{inverse} FT, necessarily remains unchanged as $p_\coh(k)\me^{-\mi k^2 t}$. On the other hand, the distribution of component $k$ values in $\abs{\Psi(x,t)}$, possessing unit phase, is the FT of a Gaussian, thereby producing a Gaussian distribution of $k$ values having standard deviation $\Dk(t)=\sigma(t)^{-1}$. However, $\Psi(x,t)$ is a pure state---\ie, a wave--packet solution of the (here, free--particle) time--dependent \schrdngr equation---while its modulus $\abs{\Psi(x,t)}$ is not. The physical difference is that, according to the Born rule, $\abs{\Psi(x,t)}$, via its square, relates directly to position \emph{measurements} of a particle in the state $\Psi(x,t)$, which brings to bear the relevance of the uncertainty principle. The state function $\Psi(x,t)$, however, does not imply a measurement outcome until, as done in the text, it is projected onto an eigenstate of a particular ``observable"; and thus its $k$--content is not subject to bounds implied by post--measurement implications of the uncertainty principle.\footnote{Of course, the state function $\Psi(x,t)$, itself, does imply the possible outcome of a position measurement because it also happens to serve as a projection (of itself) onto the eigenfunction of the position operator $\delta(x^\prime-x)$, as in the identity
\[\Dbraket{x}{\Psi(t)} = \Psi(x,t)=\Int\delta(x^\prime-x)\Psi(x^\prime,t)\, dx^{\prime}\,.\]
Here Dirac notation presents helpful abstraction. Moreover, as shown in Appendix F, the position operator is represented as
\[\xop=\Int\Dop{x}{x}{x}\, dx\] in terms of the position eigenkets $\Dket{x}$, defined so that $\Dbraket{x^\prime}{x}=\delta(x^\prime-x)$. Then it easily follows that $\xop\Dket{x}=x\Dket{x}$, and that $\Dbra{x}\xop\Dket{\Psi(t)} = x\Psi(x,t)$. It can be noted that the implied connection of $\Psi(x,t)$ to position measurements is more complicated in a relativistically invariant theory because of normalization issues; \eg, see \cite{Haag}, Sec.\ I.3.}

\section{More on wave packets}\label{Bx1}
\subsection{The Feynman integral representation}\label{Bx1.1}
It is worth pointing out that while the description of a free wave packet as a superposition of stationary plane wave states is intuitive, and by far the most commonly encountered, there are more abstract representations. For example, setting $x_0=t_0=0$ for notational purposes, we can rewrite \eqref{eq.40} in the main text---but now denoting the ``weight'' amplitude $\pco{k}$ more commonly as $w(k)$---as the inverse Fourier transform
\begin{equation}\label{eqX1}
\begin{split}
\psi(x,t) &=\int_{-\infty}^{\infty} \Bigl(w(k)\me^{-\mi(\hbar k^2/2m) t}\Bigr)\me^{\mi k x}\mdiff k/2\pi =\FTI_x\Bigl(w(k)\me^{-\mi(\hbar k^2/2m)t}\Bigr)\\
&=\int_{-\infty}^{\infty}K_0(x-y,t)\psi_0(y)\mdiff y\,,
\end{split}
\end{equation}
where $\psi_0(x)=\psi(x,0)=\FTI_x w(k)$ and the kernal---or propagator---is, for $t>0$,
\begin{equation}\label{eqX2}
K_0(x,t) = \FTI_x \me^{-\mi(\hbar k^2/2m)t} = \frac{\me^{\mi\frac{x^2}{4\beta t}}}{\sqrt{2\pi\mi\beta t}}\,,
\end{equation}
with $\beta$ as defined in Sec.\ \ref{1DWP}. The ``$0$'' subscript on $K$ here refers to the free particle Hamiltonian $\Hop=\hop$; but, as we will soon see, the basic structure \eqref{eqX1} applies to all (time--independent) $\Hop=\hop+\Vop$.\footnote{The behavior of the final equality in \eqref{eqX2} as $t\rightarrow0$ is not obvious. However, in the first equality, 
\ie, within the inverse FT, we may set $t=0$ in the integrand, at once getting $K_0(x,0)=\delta(x)$. Thus, for example, we can write $K_0(x,t)\delta(t)=K_0(x,0)\delta(t)=\delta(x)\delta(t)$, which will be applied below.} It is easy to show that, for $t>0$, $K(x,t)$ satisfies the time--dependent \schrdngr equation, so that $\psi(x,t)$ does also. Notice however, that $K(x,t)$ no longer displays a direct reference to plane waves. Indeed, in Dirac notation, the propagator is equivalent to the ``matrix element''
\begin{equation}\label{eqX10}
K(\xp-x,\tp-t)= \Dbra{\xp}\me^{-\mi/\hbar \Hop (\tp-t)}\Dket{x} = \Dbraket{\xp,\tp}{x,t}\,,
\end{equation}
where $H_\op$ is Hamiltonian operator---no longer restricted to the free particle case---and where, in the second equality the $\Dket{x,t}$ is an eigenket (also called a base ket) for the Heisenberg position operator $\xop(t)=\me^{\mi \Hop t}x_\op\me^{-\mi \Hop t}$, such that $x_{\op}(t)\Dket{x,t}=x\Dket{x,t}$. The quantity $\Dbraket{\xp,\tp}{x,t}$, sometimes referred to as a Wightman function, can be interpreted as the transition probability amplitude for ``finding'' a particle at space--time location $(\xp,\tp)$, given that it was earlier ``known'' to be at $(x,t)$.\footnote{It is a common convention in this context, especially in connection to scattering theory, to notate ``initial'' and ``final'' space--time coordinates as $(x,t)$ and $(\xp,\tp)$ respectively.}  Then we can write
\begin{equation}\label{eqX11}
\begin{split}
\psi(\xp,\tp)
&=\int_{-\infty}^{\infty}\Dbraket{\xp,\tp}{x,t}\psi(x,t)\mdiff x\,.
\end{split}
\end{equation}
This is equivalent, of course, to the textbook depiction of the general solution of the \schrdngr equation
\begin{equation}\label{eqX12.0}
\begin{split}
\Dket{\psi(\tp)}
&=\me^{-\mi/\hbar \Hop (\tp-t)}\Dket{\psi(t)}\,,
\end{split}
\end{equation}
at once satisfying $(H_\op-\mi\hbar\partial_{\tp})\Dket{\psi(\tp)}=0$. Although implicit, so far, we have not required that $\tp\ge t$. For (later) computational purposes it is helpful to make this restriction explicit by defining the (retarded) Green's function
\begin{equation}\label{eqX12}
\begin{split}
G(\xp-x,\tp-t) &= \frac{1}{\mi\hbar}\Theta(\tp-t)K(\xp-x,\tp-t)\\
&=\frac{1}{\mi\hbar}\Theta(\tp-t)\Dbraket{\xp,\tp}{x,t}\,,
\end{split}
\end{equation}
where $\Theta(t)$ is the Heaviside function. This $G(\xp-x,\tp-t)$ satisfies
\begin{equation}\label{eqX12.1}
(H_\op-\mi\hbar\partial_{\tp})G(\xp-x,\tp-t) = \delta(\tp-t)\delta(\xp-x)\,,
\end{equation}
the typical behavior for Green's functions. Here $(H_\op-\mi\hbar\partial_{\tp})\Dbraket{\xp,\tp}{x,t}$ gives $0$ on the \rhs\, as we have noted, while in \eqref{eqX12.1} the $\delta(\tp-t)$ obviously comes from $\partial_{\tp}\Theta(\tp-t)=\delta(\tp-t)$; this in turn imposes $\Dbraket{\xp,{\tp\rightarrow}t}{x,t}=\delta(\xp-x)$.\ {Given an inhomogeneous differential equation $\mathcal{L}_\op(x) f(x)=g(x)$ for a function $f(x)$ with known $g(x)$, one can express its ``special'' (or ``particular'') solution as $f(x)=\mathcal{L}_\op^{-1}(x)g(x)$ where $\mathcal{L}_\op^{-1}(x)$ is assumed to exist. But using the Green's function $G(\xp,x)$ that satisfies $\mathcal{L}_\op(\xp) G(\xp,x)=\delta(\xp-x)$, one also can write $f(\xp) = \int G(\xp,x)g(x)\,\mdiff x$, since then $g(\xp)=\mathcal{L}_\op(\xp) f(\xp)= \int\mathcal{L}_\op(\xp) G(\xp,x)g(x)\,\mdiff x = g(\zp)$, as required. Thus $G(\xp,x)$ can be viewed as an integral representative of the inverse differential operator $\mathcal{L}_\op^{-1}(\xp)$.} The Green's function itself (with a minor notational change) satisfies the integral equation

\begin{equation}\label{eqX12.2}
G(x,t)=G_0(x,t) - \int_\infty^\infty\int_\infty^\infty G_0(x-y,t-\tau) V(y)G(y,\tau)\mdiff y\mdiff\tau\,,
\end{equation}
where $G_0(x,t)$ is the free particle Green's function associated with $\hop$, since
\begin{equation}\label{eqX12.2a}
\begin{split}
(\hop(x)-&\mi\hbar\partial_{t})G(x,t) = (\hop(x)-\mi\hbar\partial_{t})G_0(x,t)\\ 
&- \int_\infty^\infty\int_\infty^\infty (\hop(x)-\mi\hbar\partial_{t})G_0(x-y,t-\tau) V(y)G(y,\tau)\mdiff y\mdiff\tau\\
&=\delta(x)\delta(t) - \int_\infty^\infty\int_\infty^\infty \delta(x-y)\delta(t-\tau) V(y)G(y,\tau)\mdiff y\mdiff\tau\\
&=\delta(x)\delta(t)-V(x)G(x,t)\,,
\end{split}
\end{equation}
giving 
$(\Hop(x)-\mi\hbar\partial_{t})G(x,t)=\delta(x)\delta(t)$, as required by \eqref{eqX12.1}.
We thus have, as shown in Note 26,
\begin{equation}\label{eqX12.4}
\begin{split}
\psi(\xp,\tp)
&=\int_{-\infty}^{\infty}G(\xp-x,\tp-t)\psi(x,t)\mdiff x\,,
\end{split}
\end{equation}
which, in this form, is usually known as the Feynman integral.\footnote{A readable mathematical account of the Feynman integral, as derived from increasingly formal points of view, is given in \cite{KellerMcLaughlin}. The Feynman integral also is called the Feynman \textsl{path} integral, but that nomenclature refers more specifically to Feynman's invention of a computational scheme for the integral involving the use of explicit particle trajectories, seemingly in contradiction to the rules of standard quantum mechanics. In essence, Feynman's trick is to observe that by casting the propagator (\ie, Green's function) in terms of an integration over all mathematically suitable particle paths, probabilistically weighted by the classical action, there is no commitment to any specific one them, and therefore no violation of the rules of SQM. In practice, depending on the Hamiltonian, the needed path integrals (actually approximated by discrete summations in the manner of numerical calculus) generally are difficult to do without invoking various approximation strategies. The Feynman integral is known to be mathematically equivalent to the Wiener integral (also known as the Wiener--Kac integral) for diffusive processes if we replace imaginary $\mi\times t$ in the quantum problem with real time $\tau$ in the diffusion problem (or vice versa); see \cite{KellerMcLaughlin} and \cite{Ito}, for example, as well as Feynman's own student oriented introduction to the notion of path integration in \cite{Feynman1}.}

Returning to 
\eqref{eqX11}, for $\tp=t$ we have  $\Dbraket{\xp,\tp}{x,t}=\Dbraket{\xp,t}{x,t}=\Dbraket{x_1}{x_0}=\delta(\xp-x)$, giving the seemingly trivial identity, $\psi(\xp,t)=\psi(\xp,t)$. The form of \eqref{eqX11}, however, might otherwise suggest that for $\tp>t_0$ $\psi(\xp,\tp)$ is a solution of the \schrdngr equation for any function $\psi(x,t)$, even one that is not itself a solution. The ``built--in'' continuity implicit in the propagator $\Dbraket{\xp,\tp}{x,t}$ as $\tp\rightarrow t$ prevents this from happening (with a notable exception to be mentioned.) Notice also that, having left behind \eqref{eqX1}, there has been no explicit reference to wave packets; however, \eqref{eqX11}, permits all solutions of the the time--dependent \schrdngr equation to be so characterized, notwithstanding that such terminology normally is reserved for applications involving some degree of spatial localization.

Let us look at specific cases of \eqref{eqX11}. Again, assume $\Hop=\hop$, and let $\psi_0(x,t)\rightarrow\psi_0(x,t|k)=\me^{\mi(k x-\hbar/2m k^2t)}$ be a given initial plane wave state. Then, with the free state propagator

\[\Dbraket{\xp,\tp}{x,t}_0 = \frac{\me^{\mi\frac{(\xp-x)^2}{4\beta(\tp-t)}}}{\sqrt{2\pi\mi\beta (\tp-t)}}\,,\]
as in \eqref{eqX2}, it follows from \eqref{eqX11}, with a little work, that

\begin{equation}\label{eqX14}
\psi(\xp,\tp) = \psi_0(\xp,\tp|k)\,;
\end{equation}
\ie, the free state kernal simply propagates free state plane waves into the future without altering their shape, as we would expect, of course. And since \eqref{eqX11} is linear in $\psi_0(x,t)$, this holds for all plane wave states comprising a free wave packet as in
\begin{equation}\label{eqX15}
\psi_0(x,t)=\int_{-\infty}^\infty w(k)\psi_0(x,t|k)\mdiff k\,.
\end{equation}
Now, however, since the various $\psi_0(x,t|k)$ move into the future at different $k$--dependent rates, the state $\psi(\xp,\tp)$ generally will not be identical to a rigidly propagated $\psi_0(x,t)$, even though the ``weight'' factor $w(k)$ defining $\psi_0(x,t)$ is fixed.\footnote{Non--spreading wave packets are briefly discussed above in Appendix \ref{sA}}

Now consider that in \eqref{eqX15} $w(k)=1$, and for convenience take $t=0$. Then (up to a constant---we will choose $\sqrt{2}$)) $\psi_0(x,0) = \sqrt{2}\delta(x)$; \ie\,, $\psi_0(x,0)$, while still a linear superposition of plane wave states, has become completely localized to $x=0$. And so, from \eqref{eqX2},

\begin{equation}\label{eqX17}
\begin{split}
\psi(\xp,\tp)
&=\Dbraket{\xp,\tp}{0,0} = \frac{\me^{\mi\frac{x^{\prime 2}}{4\beta \tp}}}{\sqrt{2\pi\mi\beta \tp}}\,.
\end{split}
\end{equation}
In this case it might be said---in a casual sense---that so many of the plane wave states comprising $\psi_0(x,0)$ so quickly move into the future at very different rates that the highly localized initial wave packet instantaneously dissipates over all space, effectively disappearing altogether. Indeed, the probability density for ``finding'' the particle at a particular space--time location is now proportional to 
\[ |\psi(\xp,\tp)|^2 = \frac{1}{2\pi\beta \tp}\,,\]
(for $\tp>0$) independent of $\xp$, behavior normally expected of independent plane waves. However, the continuity of $\psi(\xp,\tp)$ at $\tp=0$ is lost here, a consequence of the initial $\delta$--function, a spatially discontinuous function. The isolated free state kernal has no well--defined limit as $\tp\rightarrow 0$ along the real axis. However, with $\tp = -\mi\tau^\prime$, for $\tau^\prime>0$, 
\[\Dbraket{\xp,-\mi\tau^\prime}{0,0}=\frac{1}{\sqrt{2\pi\beta\tau}}\me^{\frac{-x^{\prime 2}}{4\beta\tau^\prime}}\,,\]
which is a ``real'' Gaussian having the limit $\sqrt{2}\delta(\xp)$ as $\tau^\prime\rightarrow0$ from above. Thus---with this fudge---$\psi(\xp,\tp)$ is continuous (after a fashion) at $\tp=0$ if, in the neighborhood of the $\tp$--origin, we allow $\tp\rightarrow0$ along the negative \textsl{imaginary} $\tp$--axis.

In the general case for $\Hop$ we note that since $\Dket{x,t}=\me^{\mi\Hop t/\hbar}\Dket{x}$, it follows that
\[\int_{-\infty}^\infty\mdiff x \Dketbra{x,t}{x,t} = \int_{-\infty}^\infty\mdiff x \Dketbra{x}{x} = 1\,.\] With this identity we can rewrite \eqref{eqX11} as
\begin{equation*}
\begin{split}
\psi(\xp,\tp)
&=\int_{-\infty}^{\infty}\int_{-\infty}^{\infty}\mdiff x_1\mdiff x\Dbraket{\xp,\tp}{x_1,t_1}\Dbraket{x_1,t_1}{x,t}\psi(x,t)\,.
\end{split}
\end{equation*}
Invoking this trick $N$ times gives the transition amplitude

\begin{equation}\label{eqX20}
\begin{split}
\Dbraket{\xp,\tp}{x,t}
&=\Pi_{n=1}^{N}\Bigl(\Int\mdiff x_n\Bigr)\Dbraket{\xp,\tp}{x_{N-n},t_{N-n}}\cdots\Dbraket{x_1,t_1}{x,t}\,,
\end{split}
\end{equation}
using $x_0=x$. For the case that $\Hop=H_\op^0$ nothing really is gained by this extra detail since then $\Dbraket{x_n,t_n}{x_{n-1},t_{n-1}}$ behaves in the integrals as a pseudo--Gaussian for any $n$, and the resulting integrations in \eqref{eqX20} become nested convolutions of kernals---which, of course, are then also kernels---leading us back to \eqref{eqX11}, as if $N=1$. However, because $\Hop$ contains the potential operator $V_\op$, rendering the integrations in \eqref{eqX20} no longer trivial, the chain of ``events'' in \eqref{eqX20} lead to an alternative methodology---indeed, a new quantum picture---for computing $\Dbraket{\xp,\tp}{x,t}$ when we allow $N\rightarrow\infty$. Notice, for example, that the integrand of \eqref{eqX20} can be interpreted as the joint transition probability amplitude for the temporal sequence of events that take $(x,t)$ to $(x_1,t_1)$, and then $(x_1,t_1)$ to $(x_2,t_2)$, and so on, until we reach $(\xp,\tp)$. The indicated integrations over the set $\{x_1,x_2,\cdots,x_{N-1}\}$---in the limit  $N\rightarrow\infty$---thus lead to Feynman's famous interpretation of $\abs{\Dbraket{\xp,\tp}{x,t}}^2$ as the probability density for all possible ``paths'' connecting states $\psi(x,t)$ and $\psi(\xp,\tp)$ in the presence of the Hamiltonian $\Hop$.

We conclude this section by briefly noting that in the presence of an interaction potential $V(x)$, the (retarded) Green's function acts as a solution of the integral equation

\begin{equation}\label{eqX30}
\hat{G}(t)=\hat{G}_0(t)-\int_0^t\mdiff t_1 \hat{G}_0(t-t_1)\hat{V}\hat{G}(t_1)\,.
\end{equation}
Here we have simplified notation by replacing $\tp$ with $t$ and introducing spatial operators $\hat{X}$ with the property in Dirac notation that $\Dme{x}{\hat{X}}{y}=X(x-y)$; for example, $\Dme{x}{\hat{G}(t)}{y}=G(x-y,t)$, and $\Dme{x}{\hat{V}(t)}{y}=V(x)\delta(x-y)$. Then a solution of \eqref{eqX30} as an infinite--order perturbation series follows from consecutive iterations under the integral sign, starting with $\hat{G}(t_1)\rightarrow\hat{G}_0(t_1)$, to give

\begin{equation}\label{eqX31}
\hat{G}(t)=\hat{G}_0(t)-\hat{T}\int_0^t\mdiff t_1 \hat{G}_0(t-t_1)\hat{V}\bigl[\hat{G}_0(t_1)-\int_0^t\mdiff t_2 \hat{G}_0(t_1-t_2)\hat{V}\hat{G}_0(t_2)+\cdots\bigr]\,,
\end{equation}
where $\hat{T}$ is Feynman's time--ordering operator. Unfortunately, this series generally cannot be represented in closed form (other than by replacing the bracketed quantity with $\hat{G}(t_1)$ itself, of course) without context--dependent approximations.
\section{Wave packets and second quantization}\label{WSQ}
Section\ \ref{PQFS} introduced the notion of Fock space for pure state superpositions by defining \textsl{ab initio} eigenstates of the  number operator $N_\op$, without a need---for the arguments at hand---to begin with definitions of ``creation'' and ``annihilation'' operators, in terms of which number operators usually are defined. Here we outline how such operators can be sensibly defined for wave packet states within the scope of SQM.\footnote{Historically, ordinary quantum mechanics, dealing with a fixed number of particles, is often described as ``first quantization,'' while quantum field theory and many body theory---involving transformations of particle number---generally are dubbed ``second quantization.'' Roughly speaking, first quantization can be said to refer to the elevation of (classically) physical observables (\ie, quantities existing as numerical values) to operator status in which observable values occur only by the action of these operators on state functions. Second quantization then can be said to refer to the elevation of the state functions themselves to operator status, these new operators defining ``operator--valued'' fields for generating particles (or states) from a defined \textsl{vacuum} state. Experts in these matters continue to debate the ontological status of such notions. For example, see \cite{KLW} and \cite{Kuhl}. In this appendix, we follow (more--or--less) standard language and methodology.} As in other parts of the text, We use notations similar to that in Appendix E.

\subsection{Second quantization of the canonical state pair $\Dket{k}$ and $\Dket{x}$}\label{WSQkx}

The main idea here is to generate an operator notation for the spatial representation of wave packets. To begin, we first redefine the wave number state vector $\Dket{k}$ as

\begin{subequations}\label{WSQeq.1}
\begin{equation}\label{WSQeq.1.a}
\Dket{k} = \hcd(k)\DketV\,,
\end{equation}
where $\hcd(k)$ ``creates'' $\Dket{k}$ from the \textsl{vacuum} state $\DketV$. (We specify the vacuum state as $\DketV$ to avoid possible confusion with the state $\Dket{k=0}=\Dket{0}$.) The partner operator $(\hcd(k))^\dagger = \hc(k)$ is defined to ``annihilate'' the state $\Dket{k}$---equivalently, to ``return'' it to the vacuum---in the sense that
\begin{equation}\label{WSQeq.1.b}
\hc(k)\Dket{k}=\DketV\,.
\end{equation}
\end{subequations}
Time--dependent operators can be defined for free particle states simply by taking $\hc(k)\rightarrow\hc(k)\me^{\mi\beta k^2 t}=\hc(k,t)$, so that
\begin{equation}\label{WSQeq.2}
\Dket{k,t} = \hcd(k,t)\DketV = \me^{-\mi\beta k^2 t}\Dket{k}\,.
\end{equation}
For most of what follows, we will ignore time dependence. When needed, it is easily restored for the free particle using the recipe in \eqref{WSQeq.2}. To complete the second quantization, the operators $\hcd(k)$ and $\hc(k)$, as defined for fermions, are required to be \textsl{canonical} via the anti-commutation rules
\begin{equation}\label{WSQeq.3}
\begin{split}
\{\hcd(k),\hc(k^\prime\} &= \hcd(k)\hc(k^\prime)+\hc(k^\prime)\hcd(k)\\
&=\delta(k-k^\prime)\,,
\end{split}
\end{equation}
along with $\{\hcd(k),\hcd(k^\prime)\}=\{\hc(k),\hc(k^\prime)=0\}$. The appearance of the Dirac $\delta$--function in \eqref{WSQeq.3} presents an obvious mathematical difficulty for what should be the ``easy'' case of $k^\prime=k$, but for the moment, we move on. Now, as shown in Appendix E, the free particle position eigenstate $\Dket{x}$ is given by the inverse Fourier transform (IFT) of the wave vector eigenstate $\Dket{k}$, \textsl{viz.}\ as

\begin{equation}\label{WSQeq.4}
\Dket{x}=\int \dtilk\,\me^{-\mi kx}\Dket{k}\,,
\end{equation}
with the new notation, 
\[ \dtilk=\frac{1}{2\pi}dk\,.\]
And, as also shown in Appendix E, 
\[ \Dbraket{x^\prime}{x}=\delta(x^\prime-x)\,.\]
Thus combining \eqref{WSQeq.1.a} and \eqref{WSQeq.4}, we can write
\begin{equation}\label{WSQeq.5}
\begin{split}
\Dket{x}&=\int \dtilk\,\me^{-\mi kx}\hcd(k)\DketV\\
&=\hpsid\DketV\,,
\end{split}
\end{equation}
and so, with \eqref{WSQeq.3},
\begin{equation}\label{WSQeq.6}
\begin{split}
\{\hpsid(x),\hpsi(x^\prime\}&=\int\int \dtilk \dtilk^\prime\,\me^{\mi kx-k^\prime x^\prime}\{\hcd(k),\hc(k^\prime)\}\\
&=\int\dtilk\me^{\mi k(x-x^\prime)}=
\delta(x-x^\prime)\,.
\end{split}
\end{equation}
Clearly, the $\delta(k-k^\prime)$ inherited from \eqref{WSQeq.3} was helpful here, but now we must deal with the meaning of $\{\hpsid(x),\hpsi(x)\}=\infty$. It turns out that there are two ways of removing this singularity. In the first, we project the continuum of the position variable $x$ onto a lattice of discrete positions, $\{x_j\}$, allowing us to make the replacement $\delta(x-x^\prime)\rightarrow\delta_{x_j,x_{j^\prime}}$. This approach leads to

\begin{equation*}
\{\hpsid(x_j),\hpsi(x_{j^\prime}\}=\delta_{x_j,x_{j^\prime}}\,;
\end{equation*}
or with a simpler notation, justified since $\psi(x_j)$ (in first quantization) is an eigenstate of the position operator,

\begin{equation}\label{WSQeq.7}
\{\hpsid_j,\hpsi_{j^\prime}\}=\delta_{j,j^\prime}\,,
\end{equation}

\begin{equation*}
\begin{split}
\hpsi_j\Dket{j^\prime}&=\hpsi_j\hpsid_{j^\prime}\DketV=\bigl(\delta_{j,j^\prime}-\hpsid_{j^\prime}\hpsi_j\bigr)\DketV\\
&=\delta_{j,j^\prime}\DketV\,,
\end{split}
\end{equation*}
indicating that $\hpsi_j$ annihilates the quantum state $\Dket{j}$---\ie, ``returns it'' to the vacuum---but ``sees'' only $\DketV$ when acting on $\Dket{j^\prime}$ if $j^\prime\neq j$. It also follows from \eqref{WSQeq.7} that

\begin{equation*}
\begin{split}
\hpsid_j\hpsi_j\Dket{j}&=\hpsid_j(1-\hpsid_j\hpsi_j)\DketV\\
&=\hpsid_j\DketV = 1\times\Dket{j}\,,
\end{split}
\end{equation*}
allowing the familiar identification of the number operator $\hat{N}_j$ of Fock space, as introduced in Sec.\ \ref{PQFS}, with the creation and annihilation operators of fermionic second quantification, $\hpsid_j$ and $\hpsi_j$,  as

\begin{equation}\label{WSQeq.8}
\hat{N}_j = \hpsid_j\hpsi_j\,.
\end{equation}
While this association may appear to rely on the canonical relation between the particular states, $\Dket{k}$ and $\Dket{x}$ representing eigenstates of momentum and position, the mathematics of the discrete case, in fact, applies to any first quantization relation of canonical operator pairs.

\subsection{Second quantization of a wave packet}\label{WSQkx1}

The second quantization, just derived, of the canonical pair $\Dket{k}$ and $\Dket{x}$ can be generalized to the free--particle wave packet as follows. As usual, the first quantization of the wave packet can be written as the wave function 

\begin{equation}\label{WSQeq.21}
\Psi(x,t) = \int w(k)\me^{\mi(kx - \beta k^2t)}\dtilk=\int w(k,t)\me^{\mi kx }\dtilk\,.
\end{equation}
with $ w(k,t)$ normalized to unity,

\begin{equation*}
\int |w(k,t)|^2\dtilk=\int |w(k)|^2\dtilk = 1\,.
\end{equation*}
Now define a new ``$c(k)$'' by

\begin{subequations}\label{WSQeq.22}
\begin{equation}\label{WSQeq.22a}
\hC(k,t)=w(k,t)\hc(k)\,,
\end{equation}
and a new ``$\psi(x,t)$'' by
\begin{equation}\label{WSQeq.22b}
\hPsi(x,t) = \int \hC(k,t)\me^{\mi kx}\dtilk\,,
\end{equation}
\end{subequations}
so that

\begin{equation}\label{WSQeq.25}
\hPsid(x,t) = \int\me^{-\mi kx}\hCd(k,t)\dtilk\,,
\end{equation}
and thus

\begin{equation}\label{WSQeq.26a}
\{\hPsid(x,t),\hPsi(x^\prime,t)\} = \Delta(X,T)\,,
\end{equation}
where
\begin{equation}\label{WSQeq.26b}
\Delta(X,T) = \int\me^{\mi(kX - \beta k^2T)}|w(k)|^2\dtilk\,,
\end{equation}
with $X=x-x^\prime$ and $T=t-t^\prime$. For the Gaussian wave packet in Appendix \ref{sA}, here taking normalized $w(k)$ as,

\begin{equation}
w(k) = (\frac{2\alpha}{\pi})^{\frac{1}{4}}\me^{-2\alpha(k-\kbar)^2}\,,
\end{equation}
we get
\begin{equation}\label{WPQeq.35}
\begin{split}
\Delta(X,T) &= \exp{[-\frac{\alpha(X-v_gT)^2}{2(4\alpha^2+\beta^2T^2)}]}\\
&\times\exp{[\mi\frac{4\alpha^2\kbar X+\beta(X^2/4-4\alpha^2\kbar^2)T}{4\alpha^2+\beta^2T^2}]}\,.
\end{split}
\end{equation}
where $v_g=2\beta\kbar$.
Thus, since, in general, $\Delta(0,0)=0$, we have that

\begin{equation}\label{WPQeq.40}
\hat{N}(x,t)=\hPsid(x,t)\hPsi(x,t)
\end{equation}
acts as a proper number operator on the state
\begin{equation}\label{WPQeq41}
\Dket{\Psi(x,t)}=\hPsid(x,t)\DketV\,,
\end{equation}
since
\begin{equation}\label{WPQeq.42}
\hat{N}(x,t)\Dket{\Psi(x,t)}=1\times\Dket{\Psi(x,t)}\,,
\end{equation}
independent of $(x,t)$ as required.
It is important to emphasize, however, that $\hPsid(x,t)$ creates the \textsl{state vector} $\Dket{\Psi(x,t)}$ at coordinate $(x,t)$. Let us now call \textsl{this} $(x,t)$, $(x_0,t_0)$. The associated \textsl{wave function}, $\Psi(x-x_0,t-t_0)$ is then given by the Dirac projection

\begin{equation}\label{WPQeq.43}
\Psi(x-x_0,t-t_0)=\Dbraket{x-x_0,t-t_0}{\Psi(0,0)}\,.
\end{equation}

It is worth noting that localized wave packets representing ``smeared out'' coordinate representations of state vectors were originally introduced into quantum field theory by Wightman \cite{Wightman} in order to remove the troublesome $\delta$--function singularities associated with position eigenstates $\Dket{x,t}$. He referred to these as \textsl{test functions} rather than wave packets, apparently giving emphasis to their mathematical role over physical interpretation. Test functions, however, brought along difficulties of their own into the mathematical development of the theory, leading to what came to be known as \textsl{axiomatic} QFT. See \cite{KLW}, for example.

\section{Proof of a certain property of statistical operators}\label{sE}

First, some preliminaries. Let 

\begin{equation}\label{eq.WPTA1}
\Dket{\psi^{(\chi)}(t)} = \sum_\nu a_\nu^{(\chi)}\Dket{\chi_\nu(t)}
\end{equation}
be a solution of the \schrdngr equation for Hamiltonian $H$ represented in the orthonormal basis $\Dket{\chi_\nu(t)}$, where $\Dket{\chi_\nu(t)}=\me^{-\mi H t}\Dket{\chi_\nu}$ and 
$\Dket{\chi_\nu}$ is an eigenket of an Hermitian operator $\chi$ belonging to eigenvalue $\chi_\nu$, so that $\chi\Dket{\chi_\nu}=\chi_\nu\Dket{\chi_\nu}$.\footnote{Here we use a simplified notation we hope is clear.} But we can also express every $\Dket{\chi_\nu}$ in the energy basis $\Dket{\phi_n^{(E)}}$, where $H\Dket{\phi_n^{(E)}}=E_n\Dket{\phi_n^{(E)}}$, giving $\Dket{\chi_\nu}=\sum_n b_{\nu n}^{(\chi,E)}\Dket{\phi_n^{(E)}}$ with $\Dket{\chi_\nu(t)} = \sum_n b_{\nu n}^{(\chi,E))} \me^{-\mi E_n t}\Dket{\phi_n^{(E)}}$, and easily arriving at

\begin{equation}\label{eq.WPTA2}
\Dket{\psi^{(\chi)}(t)} = \sum_{n} c_n^{(\chi, E)}\me^{-\mi E_n t}\Dket{\phi_n^{(E)}}\,,
\end{equation}
where
\begin{equation}\label{eq.WPTA3}
c_n^{(\chi, E)} = \sum_\nu^{(\chi)}c_{\nu n}^{(\chi, E)}\,,
\end{equation}
with $c_{\nu n}^{(\chi, E)}= a_\nu^{(\chi)} b_{\nu n}^{(\chi,E)}$.
Since $\Dket{\psi^{(\chi)}(t)}$ is by definition a \textsl{pure state}, its representation as a statistical operator (density matrix),

\begin{equation}\label{eq.WPTA4}
\rho_\pure^{(\chi)}(t) = \Dket{\psi^{(\chi)}(t)}\Dbra{\psi^{(\chi)}(t)}\,,
\end{equation}
also is a pure state, albeit now in the form of a projection operator. Writing out $\rho_\pure^{(\chi)}(t)$ in the representation of \eqref{eq.WPTA2} one gets, with a bit of rearranging, the sum of two terms,

\begin{equation}\label{eq.WPTA5}
\begin{split}
\rho_\pure^{(\chi)}(t) &= \sum_{n} \abs{c_{n}}^2\Dketbra{\phi_n^{(E)}}{\phi_n^{(E)}}\\
 &+\sum_{\nu,\nu^\prime}\sum_{n,m\ne n} c_{\nu n}^{(\chi, E)} c_{\nu^\prime m}^*{(\chi, E)}\me^{-\mi (E_n-E_m) t}\Dketbra{\phi_n^{(E)}}{\phi_m^{(E)}}\,.
\end{split}
\end{equation}

Now taking up an idea introduced by Ballentine,\footnote{\cite{Ballentine}, Sec. 9.4, pp. 238-241.} for the case of free particles, we consider the notion of the time--average of a pure--state statistical operator defined as

\begin{equation}\label{eq.WPTA6}
\bigl<\rho_\pure^{(\chi)}(t)\bigr>_t = \lim_{T\rightarrow\infty}\frac{1}{T}\int_{T/2}^{T/2}\rho_\pure(t)dt\,.
\end{equation}
Applying this average to $\rho_\pure^{(\chi)}(t)$ in \eqref{eq.WPTA5}, we notice that the first term is unaffected since it is independent of $t$, while in the second term the time integral affects only the propagation factor (taking $E_n-E_m=\Delta_{m,n}$),
\begin{equation}\label{eq.WPTA7}\begin{split}
\lim_{T\rightarrow\infty}\frac{1}{T}\int_{T/2}^{T/2}\me^{\mi\Delta_{m,n} t}dt =
\lim_{T\rightarrow\infty}\sinc{\bigl(\Delta_{m,n}T/2\bigr)}=\delta_{m,n}\,,
\end{split}
\end{equation}
where $\sinc{x}=\sin{x}/x$ and $\delta_{m,n}$ is the Kronecker delta: $\delta_{n,n}=1$, $\delta_{m\ne n, n}=0$.\footnote{We use $\sinc{(0)}=1$ when  $\Delta_{n,n}=0$, and $\sinc{(\infty)}=0$ when $\Delta_{m,n}\ne 0$, assuming the absence of degeneracy.} Thus, since $m\ne n$ in the second term, it vanishes at once, and we have

\begin{equation}\label{eq.WPTA8}
\bigl<\rho_\pure^{(\chi)}(t)\bigr>_t = \sum_n \abs{c_n^{(\chi,E)}}^2\Dketbra{\phi_n^{(E)}}{\phi_n^{(E)}}\,,
\end{equation}
Or with $\rho_\mix^{(E)}=\sum_n \abs{c_n}^2\Dketbra{\phi_n^{(E)}}{\phi_n^{(E)}}$ for appropriate $c_n$,
\begin{equation}\label{eq.WPTA9}
\bigl<\rho_\pure(t)\bigr>_t = \rho_\mix^{(E)}
\end{equation}
for \textsl{any} $\rho_\pure(t)$. We bear in mind, however, recalling \eqref{eq.WPTA3}, that the probabilities $p_n(E) = \abs{c_n^{(\chi,E)}}^2$ associated with the derived mixed state are not, in general, the same probabilities $p_n(\chi) = \abs{a_\nu^{(\chi)}}^2$ associated with the weights of superposition in the originally constructed pure state $\Dket{\chi_\nu(t)}$.

On the face of it, \eqref{eq.WPTA9} may not be surprising, considering that any superposition of pure state projection operators is---essentially by definition---a mixed state, and time averaging can casually (at least) be thought of in such terms. The real question, however, is: Why compute $\bigl<\rho_\pure(t)\bigr>_t$? Ballentine's answer appears to us to beg the question by its assumption that in scattering experiments the statistical operator representing the incident beam must describe a stationary state requiring $\partial\rho/\partial t=0$, so that $\rho$ and $H$ ``possess a complete set of common eigenvectors [of free particles].'' Then, the argument goes, the statistical matrix representing a single incident wave packet must be made consistent with this property, which can be achieved by averaging it over all time.\footnote{It is also stated (\cite{Ballentine}, top of p.\ 240) that ``[a]ll observable quantities...can  be calculated from the [incident] state function $\rho(x,x^\prime)[=\Dme{x}{\rho_\mix^{(\mathrm{inc},E)}}{x^\prime}$],'' which would seem to eliminate time--sensitive scattering effects}

Such an explanation leaves open the question of whether the time--average of $\rho_\pure(t)$ is physically meaningful.\footnote{Specifically, for fixed $t$, Ballentine averages over a $t_0$, which he associates with the creation time of a wave packet. However, the time dependence of his model is defined solely by the temporal difference $t-t_0$, so that the average could just as well be over all $t$ for fixed $t_0$ or over the time difference $\tau=t-t_0$, yielding identical results corresponding to our \eqref{eq.WPTA9} and thus introducing an ambiguity into any interpretation of the average.} Our view is that, in the present context, it is not. An unweighted average of any $\rho_\pure(t)$ over all $t$ generates unphysical superpositions of wave packet states that can be interpreted as moving both forward and backward, while---for \schrdngr wave packets---broadening in both directions, thereby eliminating meaningful propagation of wave packets toward a fixed target.\footnote{In this regard it is interesting to recall Note 5 of Sec.\ IB.} The pure state generates all statistically possible detector responses (each an isolated ``click'' or ``spot'') of individual wave packets from a given class (see Sec.\ IIc) in a given setup at a given time; and, as shown in Sec.\ II, wave packet scattering is strongly $t$--dependent over a range of times that in principle can be controlled.

It could be said, however, that \eqref{eq.WPTA9} does reveal a mathematical property of the pure state statistical operator $\rho_\pure(t)=\Dketbra{\Psi(t)}{\Psi(t)}$, albeit one of passing interest. From the definition of the reflection amplitude given in Sec.\ IIC (slightly simplified) as $r(k,t)=\Dbraket{-k,t}{\Psi(t)}$, we easily have that

\begin{equation}\label{eq.WPTA10}
\begin{split}
&\lim_{t\rightarrow\infty}\Dme{-k,t}{\rho_\pure(t)}{-k,t}=\abs{\Dbraket{k}{\Psi(\infty)}}^2\\
&= \abs{r(k,\infty}^2 = R_\coh(k) = P_\coh(k)R_\pw(k)\,,
\end{split}
\end{equation}
which would be consistent with \eqref{eq.WPTA9} were \textsl{it} written as

\begin{equation}\label{eq.WPTA11}
\begin{split}
&\Dme{-k}{\bigl<\rho_\pure(\infty)\bigr>_t}{-k}=\abs{\Dbraket{k}{\Psi(\infty)}}^2\\
&= P_\coh(k)R_\pw(k)\,.
\end{split}
\end{equation}
Roughly put, in the limit of infinite $t$, the time average of $\rho_\pure(t)$ over an ``infinite--t'' range effectively behaves as an average over a $t$--independent (``already converged'') statistical operator. Thus, while \eqref{eq.WPTA9} is true---as a  formality---it is physically relevant only to the long--time limit, where in essence it becomes tautological.

\section{The transfer matrix}\label{sB} 

Here we review the powerful transfer matrix method (TMM) for solving the 1--D plane wave stationary state reflection/transmission problem for an arbitrarily shaped barrier of length $L$. These solutions provide a convenient basis for constructing the corresponding wave packet solutions for the given barrier.

As in Sec.\ \ref{1DWP1} it is convenient to first break the $x$--axis into three disjoint regions: $\rgn{I}$ for $x<0$, $\rgn{II}$ for $0\le x\le L$, and $\rgn{III}$ for $x>L$. The barrier is thus supported on region $\rgn{II}$, and we will allow $\rgn{III}$ to be filled with a constant $q=q_b$. Thus region $\rgn{I}$ holds the incident and reflected contributions to $\psi(k,x)$ while $\rgn{III}$ holds the transmitted contribution. The generic physical solutions in \rgn{I} and \rgn{III}, the ones of usual direct concern, as earlier,

\begin{equation}\label{c6eq1002}
    \psi(k,x) = \begin{cases}
    \psi_\rgn{I}(k,x)=e^{in_f(k)kx}+r_\pw(k)e^{-n_f(k)kx} \ \ \mathrm{and}\\
    \psi_\rgn{III}(k,x)=t_\pw(k)e^{in_b(k)kx}
    \end{cases}
\end{equation}

\noindent with $r_\pw(k)$ and $t_\pw(k))$ to be determined, and where (for additional generality) we have allowed for both regions \rgn{I} and \rgn{III} to be treated as non--vacuum ``fronting'' and ``backing,'' respectively, with $q_\rgn{I}(x)=q_f$ and index of refraction $n_f(k)$, and with $q_\rgn{III}(x)=q_b$ and index of refraction $n_b(k)$. Here again, as in Sec.\ \ref{1DWP1}, we begin with standard phase definitions and introduce the appropriate phases for the wave packet construction when needed. In the often discussed Fresnel problem  (\ie, no barrier present, equivalent either to the absence of a region \rgn{II} or to $L=0$ in $\rgn{II}$, but a step--like change from fronting to backing at $x=0$) we could apply the required wave function continuity conditions directly to these piecewise solution elements, since they share a common boundary. In general, however, $\psi_\rgn{I}(k,x)$ and $\psi_\rgn{III}(k,x)$ each abut the intervening---but as yet unknown---solution in region \rgn{II}.  On the other hand, we know that whatever form $\psi_\rgn{II}(k,x)$ takes, both it and its first derivative must be continuous with $\psi_\rgn{I}(k,x)$ at $x=0$ and with $\psi_\rgn{III}(k,x)$ at $x=L$. The ``trick," then, is to transfer the explicit continuity of $\psi_\rgn{II}(k,x)$ across region \rgn{II}, where continuity with $\psi_\rgn{III}(k,x)$ also can be made explicit. This can be accomplished by first incorporating $\psi(k,x)$ and its first derivative with respect to $x$, $\psi^\prime(k,x)$, into the column vector

\begin{equation}\label{c6eq1003}
    \chi(k,x) = \begin{pmatrix}
    \psi(k,x)\\k^{-1}\psi^\prime(k,x)
    \end{pmatrix}
\end{equation}

\noindent at all $x$. The factor $k^{-1}$ in the bottom element is not required but gives both elements of $\chi(k,x)$ the common dimension of inverse length, which ultimately leads to neater looking formulas. By rough analogy with classical mechanics, once may think of the seemingly redundant representation $\psi(k,x)\to\chi(k,x)$ as a transformation from ``configuration space" to ``phase space," in which the two required continuity conditions on $\psi(k,x)$ are conveniently compressed into a single condition on the state $\chi(k,x)$.

\noindent Now introduce a 2X2 matrix $M(k,x)$

\begin{equation}\label{c6eq1004}
    M(k,x) = \begin{pmatrix}
    A(k,x)&B(k,x)\\C(k,x)&D(k,x)
    \end{pmatrix}\,,
\end{equation}

\noindent asserting the property

\begin{equation}\label{c6eq1005}
    \chi(k,x) =  M(k,x)\chi(k,0)\,,
\end{equation}

\noindent and thereby ``transferring'' $\chi(k,0)=\chi_\rgn{I}(k,0)$ across region \rgn{II} to any desired $x$, up to and including $x=L$, where 
\begin{equation}\label{c6eq1005.1}
    \chi(k,L) = \chi_\rgn{III}(k,L) =  M(k,L)\chi_\rgn{I}(k,0)\,,
\end{equation} 
and with consistency at $x=0$ requiring

\begin{equation}\label{c6eq1006}
    M(k,0) =  1\,,
\end{equation}

\noindent where ``1" stands for the unit matrix. Since in the present context $k$ acts as a constant label, and in order to make subsequent equations and formulas easier to read, we now suppress the $k$ argument until it is needed. In fact, we will sometimes suppress both $(x,t)$ arguments when there is no likelihood of confusion.

The wave equation for $\psi(k,x)\rightarrow\psi(x)$ must then lead to an evolution equation for $M(x)$. To find it, first, differentiate both sides of \eqref{c6eq1005} with respect to $x$,

\begin{equation}\label{c6eq1007}
    \chi^\prime(x) =  M^\prime(x)\chi(0)\,.
\end{equation}

\noindent Then use the stationary state \schrdngr equation,

\begin{equation}\label{c6eq1000}
    -\psi^{\dprime}(x) + q(x)\psi(x)=k^2\psi(x)\,,
\end{equation} 

to give

\begin{equation}\label{c6eq1008}
    \chi^\prime(x) = \begin{pmatrix}
      \psi^\prime(x)\\k^{-1}\psi^{\dprime}(x)
    \end{pmatrix}=
    \begin{pmatrix}
      \psi^\prime(x)\\ [k^{-1}q(x)-k]\psi(x)
       \end{pmatrix}\,,
\end{equation}

\noindent which is easily rearranged as

\begin{equation}\label{c6eq1009}
    \chi^\prime(x) =  \Gamma(x)\chi(x)\,,
\end{equation}

\noindent with

\begin{equation}\label{c6eq1010}
    \Gamma(x) = \begin{pmatrix}
    0&k\\k^{-1}q(x)-k&0
    \end{pmatrix}\,.
\end{equation}

\noindent A first--order differential equation for $M(x)$ then follows by substituting \eqref{c6eq1009} into \eqref{c6eq1007} and using \eqref{c6eq1005} to eliminate $\chi(x)$, leading to

\begin{equation}\label{c6eq1011}
    M^\prime(x) =  \Gamma(x)M(x)\,,
\end{equation}

\noindent with initial condition $M(0)=1$. This has unique solutions for given $q(x)$ (and $k$). Expansion of \eqref{c6eq1011} produces a set of coupled first--order equations for the elements of $M(x)$; \viz

\begin{equation}\label{c6eq1012}
\begin{split}
A^\prime(x)&=kC(x)\,, \\
B^\prime(x)&=kD(x)\,, \\
C^\prime(x)&=[k^{-1}q(x)-k]A(x)\,, \\
D^\prime(x)&=[k^{-1}q(x)-k]B(x)\,,
\end{split}
\end{equation}

\noindent with initial data

\begin{equation}\label{c6eq1013}
\begin{split}
A(0)&=1\,, \\
B(0)&=0\,, \\
C(0)&=0\,, \\
D(0)&=1\,.
\end{split}
\end{equation}

With these formulas in hand, we can deduce several properties of $M(x)$ that pertain to any $q(x)$ in region \rgn{II}. For example, differentiating the determinant $|M(x)|$ and using \eqref{c6eq1012} gives

\begin{equation}\label{c6eq1020}
\begin{split}
\frac{\mathrm{d}}{\mathrm{d}x}|M(x)| &= A^\prime D + A D^\prime - B^\prime C - B C^\prime\\
&=0
\end{split}
\end{equation}
independent of $x$. Thus, since, from \eqref{c6eq1013}, $|M(0)|=1$, it follows that $|M(x)|=1$, independent of $x$; that is to say, $M(x)$ is \emph{unimodular}.\footnote{Because of this, $M(x)$ also is \emph{simplectic}. This means that given the matrix
\begin{equation*}
\mathcal{I} = \begin{pmatrix}
    0 & 1\\-1 & 0
    \end{pmatrix}\,,
\end{equation*}
then
\begin{equation*}
M^\top(x)\mathcal{I}M(x)=\mathcal{I}\,.
\end{equation*}
The ultimate implications of this abstruse property are subtle, but in essence lead, via conservation of current (or ``Hamiltonian flow'' in classical terms), to the existence of a unique $M(x)$ satisfying $\chi(x)=M(x-x^\prime)\chi(x^\prime)$ for any $q(x)$ in region \rgn{II}.
}

For a piecewise form of $q(x)$, as given by \eqref{eq.0}, one can show that for $x_J< x<x_{J-1}$ (with $x_J=L$ and $x_0=0$ as earlier), $M(x)$ in the last segment can be written as the matrix decomposition

\begin{equation}\label{c6eq1021}
\begin{split}
M(x) &= M_J(x-x_{J-1})\cdots M_j(x_{j-1}-x_{j-2})\cdots M_2(x_2-x_1)M_1(x_1)\\
&= M_J(x-x_{J-1})\cdots M_j(\Delta L_{j-1})\cdots M_2(\Delta L_2)M_1(\Delta L_1)\,,
\end{split}
\end{equation}
where $\Delta L_j=x_j-x_{j-1}$ and each $M_j(x-x_{j-1})$ depends only on $q_n(x)$. A rigorous proof of \eqref{c6eq1021} is somewhat busy, but the result is consistent with easy induction.\footnote{\Eg, with $x_2\ge x>x_1$, start with $\chi(x)=M(x)\chi(0)=M_2(x-x_1)\chi(x_1)=M_2(x-x_1)M_1(x_1-0)\chi(0)$, and so on.} Even more ``mathematically,'' integrating \eqref{c6eq1009} gives, for $x\ge0$,

\begin{equation}\label{c6eq1025}
\begin{split}
M(x) = 1 + \int_0^x dx^\prime\Gamma(x^\prime)M(x^\prime)\,,
\end{split}
\end{equation}
since $M(0)=1$, which in turn can be solved by successive iteration to produce the series

\begin{equation}\label{c6eq1026}
\begin{split}
M(x) = 1 + \int_0^x dx^\prime\Gamma(x^\prime)+\int_0^x dx^{\prime}\Gamma(x^{\prime})\int_0^{x^\prime} dx^{\dprime} \Gamma(x^{\dprime})+
\cdots\,.
\end{split}
\end{equation}
Now using Feynman's time--ordering trick---here as $x$--ordering---this series can be rewritten symmetrically as

\begin{equation}\label{c6eq1027}
\begin{split}
M(x) = 1 + X\sum_{n=1}^\infty\frac{1}{n!}\int_0^x\cdots\int_0^x dx_n\cdots dx_1\Gamma(x_n)\cdots\Gamma(x_1)\,,
\end{split}
\end{equation}
where $X$ is the $x$--ordering operator defined by

\begin{equation}\label{c6eq1028}
\begin{split}
X\{\Gamma(x^\prime)\Gamma(x^{\dprime})\}&=X\{\Gamma(x^{\dprime})\Gamma(x^{\prime})\}\\&=
\begin{cases}
\Gamma(x^\prime)\Gamma(x^{\dprime}) & \text{if $x^\prime>x^{\dprime}$}\\
\Gamma(x^{\dprime})\Gamma(x^\prime) & \text{if $x^{\dprime}>x^{\prime}$}
\end{cases}
\end{split}\,.
\end{equation}
The imposed ordering in \eqref{c6eq1027} is required, since in general $\Gamma(x^{\prime})$ and $\Gamma(x^{\dprime})$ commute only if $q(x^{\dprime})=q(x^{\prime})$, while the factorial in the summation corrects (before the fact) for the over--counting of all the permutations that $X$ reorders (after the fact). The (unordered) summation is, of course, just the Maclaurin series for an exponential; \ie,

\begin{subequations}\label{c6eq1029}
\begin{equation}\label{c6eq1029a}
M(x) = X\me^{\int_0^x dx^\prime\Gamma(x^\prime)}
\end{equation}
for any $q(x^\prime)$ on the interval $\{0,x\}$. And therefore,
\begin{equation}\label{c6eq1029b}
\chi(x) = X\me^{\int_0^x dx^\prime\Gamma(x^\prime)}\chi(0)\,,
\end{equation}
\end{subequations}
for $0\le x \le L$.

Now given $M(x)$, and using \eqref{c6eq1002}, we have

\begin{equation}\label{c6eq1040}
    \chi(L) =
    M(L)\chi(0)\,,
\end{equation}
so that
\begin{equation}\label{c6eq1041}
    t_\pw\!b\,\me^{ikL} 
    \begin{pmatrix}
    1\\ \mi
    \end{pmatrix} =
    \begin{pmatrix}
    A & B\\C & D
    \end{pmatrix}
   \begin{pmatrix}
   1+r_\pw\\ \mi b (1-r_\pw)\,,
   \end{pmatrix}
\end{equation}
with $A(L) = A$, etc; and so
\begin{subequations}\label{c6eq1042}
\begin{equation}\label{c6eq1042a}
r_\pw = \frac{fbB+C+\mi (fD-bA)}{fbB-C+\mi (fD+bA)}
\end{equation}
and
\begin{equation}\label{c6eq1042b}
t_\pw = \frac{2\mi\!f\,\me^{-\mi bkL}}{fbB-C+\mi (fD+bA)}\,,
\end{equation}
\end{subequations}
where for continued notational simplicity we now use $n_f=f$ and $n_b=b$.\footnote{So for vacuum fronting and backing,
\begin{equation*}
r_\pw = \frac{B+C+\mi (D-A)}{B-C+\mi (D+A)}
\end{equation*}
and
\begin{equation*}
t_\pw = \frac{2\mi\me^{-\mi kL}}{B-C+\mi (D+A)}\,.
\end{equation*}
}
It is valuable to note that in taking advantage of the unimodular property of $M$, we can---after some algebra---also find that
\begin{equation}\label{c6eq1043}
r_\pw^{fb} = \frac{\alpha^{fb}-\beta^{fb}+2\mi\gamma^{fb}}{\alpha^{fb}+\beta^{fb}+2}\,,
\end{equation}
with
\begin{subequations}\label{c6eq1043.2}
\begin{equation}
\alpha^{fb} = f^{-1}\bigl(bA^2+b^{-1}C^2\bigl)\,,
\end{equation}
\begin{equation}
\beta^{fb} = f\bigl(bB^2+b^{-1}D^2\bigl)\,,
\end{equation} and
\begin{equation}
\gamma^{fb} = b(AB+CD) = \sqrt{\alpha^{fb}\beta^{fb}-1}\,.
\end{equation} 
\end{subequations}
And then
\begin{equation}\label{c6eq1043.5}
\abs{r_\pw^{fb}}^2=R_\pw^{fb} = \frac{\Sigma^{fb}-2}{\Sigma^{fb}+2}
\end{equation}
with
\begin{equation}\label{c6eq1043.6}
\Sigma^{fb} = \alpha^{fb} + \beta^{fb}\,.
\end{equation}
If we introduce a new unimodular matrix as

\begin{equation}\label{c6eq1094}
\begin{split}
    \tilde{M}^{fb}&\equiv
    \begin{pmatrix}
    \sqrt{b}&0\\0&1/\sqrt{b}
    \end{pmatrix}
    \begin{pmatrix}
    A&B\\C&D
    \end{pmatrix}
    \begin{pmatrix}
    1/\sqrt{f}&0\\0&\sqrt{f}
    \end{pmatrix}\\
    &= \begin{pmatrix}
    \Tilde{A}^{fb}&\Tilde{B}^{fb}\\\Tilde{C}^{fb}&\Tilde{D}^{fb}
    \end{pmatrix}\,,
\end{split}
\end{equation}

\noindent then

\begin{equation}\label{c6eq1095}
(\tilde{M}^{fb})^\top \tilde{M}^{fb} =
\begin{pmatrix}
\alpha^{fb}&\gamma^{fb}\\ \gamma^{fb}&\beta^{fb}
\end{pmatrix}\,,
\end{equation}

\noindent so that

\begin{equation}\label{c6eq1096}
\Sigma^{fb}=\trace\bigl[\bigl(\tilde{M}^{fb})^\top \tilde{M}^{fb}\bigr]\,.
\end{equation}
Now for any real--valued square matrix $M$, the Hilbert--Schmidt (or Frobenius) norm is defined by $\abs{\abs{M}}_\mathrm{HS}=\sqrt{\trace\bigl(M^\top M\bigr)}$, which is the square--root of the sum of squares of its elements. Thus $\Sigma_\pw^{fb}=\abs{\abs{\tilde{M}^{fb}}}_\mathrm{HS}^2$, so \eqref{c6eq1043.5} can be recast as\footnote{Formally,

\begin{equation*}
    \tilde{M}^{fb} = \begin{pmatrix}\tilde{A}&\tilde{B}\\\tilde{C}&\tilde{D}\end{pmatrix}
\end{equation*}

\noindent may be considered as the transfer matrix for the faux free--film problem in which $\tilde{\psi}_\rgn{I} = e^{ik_xx}+\tilde{r}e^{ik_xx}$ and $\tilde{\psi}_\rgn{III} = \tilde{t}e^{ik_xx}$, so that

\begin{equation*}
    \begin{pmatrix}
    1\\i
    \end{pmatrix}\tilde{t}e^{ik_xL} =
    \begin{pmatrix}\tilde{A}^{fb}&\tilde{B}^{fb}\\\tilde{C}^{fb}&\tilde{D}^{fb}\end{pmatrix}
     \begin{pmatrix}
    1+\tilde{r}\\i(1-\tilde{r})
    \end{pmatrix}\,,
\end{equation*}
for the exact $\tilde{r}=r_\pw^{fb}$ and $\tilde{t}=t_\pw^{fb}$.}

\begin{equation*}
    R_\pw^{fb} = \frac{\abs{\abs{\tilde{M}^{fb}}}_\mathrm{HS}^2 - 2}{\abs{\abs{\tilde{M}^{fb}}}_\mathrm{HS}^2+2}\,.
\end{equation*}

The transmission amplitude in \eqref{c6eq1042b} does not reduce to a function only of $\alpha^{fb}$, $\beta^{fb}$, and $\gamma^{fb}$, but for $T_\pw^{fb}=\abs{t_\pw^{fb}}^2$ we have

\begin{equation*}
T_\pw^{fb}=\frac{4f^2}{(fbB-C)^2+(fD+bA)^2}=\frac{4fb^{-1}}{\alpha^{fb}+\beta^{fb}+2}=fb^{-1}(1-R_\pw^{fb})\,,
\end{equation*}

\noindent or

\begin{equation}\label{c6eq1098}
R_\pw^{fb}+bf^{-1}T_\pw^{fb}=1\,.
\end{equation}

As a computational matter, $q_\rgn{II}(x)$ is almost always modeled by piecewise continuous rectangular ``bins'' so that $q_{x_j,x_{j-1}}=q_j$ for $x_{j}\ge x \ge x_{j-1}$, and with $x_j-x_{j-1}=L_j$. Then $M_j$ (in the $j$--th bin ) takes the simple form

\begin{equation}\label{c6eq1045}
M_j = \begin{pmatrix}
\cos(n_jkL_j) & n_j^{-1}\sin(n_jkL_j)\\
-n_j\sin(n_jkL_j) & \cos(n_jkL_j)
\end{pmatrix}\,.
\end{equation}

As an application of TMM to complement the earlier treatment of the Fresnel problem, we now set up some ``code'' for the case that region \rgn{II} holds a piecewise continuous $q(x)$ consisting of three
bins of constant $q(x)= q_j$, for $j=1,2,3$. Thus we begin with

\begin{equation}\label{c6eq1050}
M(x) = 
\begin{cases}
M_1(x) & \text{for $x\in\{0,x_1\}$}\\
M_2(x-x_1)M_1 & \text{for $x\in\{x_1,x_2\}$}\\
M_3(x-x_2)M_2M_1 & \text{for $x\in\{x_2,x_3\}$}
\end{cases}\,,
\end{equation}
where now: each $x_j$ is a sum of bin widths, $x_j=\sum_{j^\prime=1}^j L_{j^\prime}$; and each $M_j(x-x_{j-1})$ is given by the \emph{rhs} of \eqref{c6eq1045} with $L_j$ replaced by $x-x_{j-1}$. We notate the symbolic matrix elements $A_j$, \emph{etc.}, in like fashion. We now can write

\begin{equation}\label{c6eq1052}
\chi(x) = 
\begin{cases}
\chi_\rgn{I}(x) & \text{for $x<0$}\\
\chi_1(x)=M_1(x)\chi_\rgn{I}(0) & \text{for $x\in\{0,x_1\}$}\\
\chi_2(x)=M_2(x-x_1)\chi_1(x_1) & \text{for $x\in\{x_1,x_2\}$}\\
\chi_3(x)=M_3(x-x_2)\chi_2(x_2) & \text{for $x\in\{x_2,x_3\}$}\\
\chi_\rgn{III}(x)=\chi_3(L) & \text{for $x>L$}
\end{cases}\,,
\end{equation}
or more simply,
\begin{equation}\label{c6eq1052.5}
\chi(x) = 
\begin{cases}
\chi_\rgn{I}(x) & \text{for $x<0$}\\
\chi_\rgn{II}(x)=M(x)\chi_\rgn{I}(0) & \text{for $x\in\{0,x_3\}$}\\
\chi_\rgn{III}(x)=\chi_3(L) & \text{for $x>L$}
\end{cases}\,,
\end{equation}
where $\chi_\rgn{II}(x)=\sum_{j=1,3}\chi_j(x)$, since now $M(x)$ is known from \eqref{c6eq1052}.
Our goal, of course, is to use the plane wave solution for $\psi(x|k)$ to produce a wave packet $\Psi(x)$; \ie, to take $\psi(x|k)\rightarrow\Psi(x)$ via an integral transformation $\mathcal{P}_k$ such that
$\Psi(x) = \mathcal{P}_k\psi(x|k)$. With TMM this is accomplished by applying $\mathcal{P}_k$ to the vector $\chi(x|k)$ to give 

\begin{equation}\label{c6eq1053}
\mathcal{P}_k\chi_(x|k)=\mathcal{P}_k
\begin{pmatrix}
\psi(x|k)\\
k^{-1}\psi^\prime(x|k)
\end{pmatrix} = 
\begin{pmatrix}
\mathcal{P}_k\psi(x|k)\\
\mathcal{P}_k k^{-1}\psi^\prime(x|k)
\end{pmatrix}
\,,
\end{equation}
and then discarding the lower of the two elements.

Perhaps needless to say, the example here of a 3--component rectangular barrier in region $\rgn{II}$ is easily extended to an $n$--component barrier for any integer $n$. Less obvious, perhaps, is that more can be said about the mathematical structure of TMM. See \cite{MB10} and \cite{MB11}, for example, for other useful formulas; also see \cite{MB12} for the extension of TMM to particles with spin.

\section{Outline of Dirac notation}\label{sF}

In Dirac notation a quantum state is symbolically represented by a ``vector'' denoted by the symbol $\Dket{\psi}$, called a \emph{ket}. The scalar (or inner) product of two kets is indicated by the symbol $\Dbraket{\psi_2}{\psi_1}$, where $\Dbra{\psi}$, standing alone, is called the \emph{bra} or ``dual'' of the ket $\Dket{\psi}$.\footnote{In the formal mathematics of vector and function spaces, duals in general are constructs defined for the purpose of mapping vectors onto scalars---or functions onto numbers; roughly speaking, the dual of an object provides something for any other similar object to be projected on to. In function space, the dual of $\psi(x)$ is its complex conjugate $\psi^*(x)$. Be alert, however, to variations in the definitions of duals. In some conventions the dual of a given $\Dket{\psi}$ is taken to be the projection of $\Dket{\psi}$ onto a specified basis---\ie, the set of scalars $\{\Dbraket{\phi_n}{\psi}\}$. These dual scalar ``spaces'' then have the same properties of linearity and completeness as the originating vector spaces.} To make contact with function space, the need arises to define a transformation of the ket $\Dket{\psi}$ to the function $\psi(x)$ it stands for. The Dirac method for this is to introduce a complete set of orthogonal position kets $\Dket{x}$, and their duals $\Dbra{x}$, such that
\begin{equation}\label{eqA.1}
\Dbraket{x}{\psi}=\psi(x)\,,
\end{equation}
for any $\Dket{\psi}$ and having the properties
\begin{equation}\label{eqA.2}
\Dbraket{x}{x^\prime}=\delta(x^\prime-x)\,,
\end{equation} where $\delta(x)$ is the Dirac delta function, and
\begin{equation}\label{eqA.3}
\int\Dketbra{x}{x} \mdiff x = 1\,
\end{equation}
is the unit operator. Then
\begin{equation*}
\begin{split}
\Dbraket{\psi_2}{\psi_1} &= \Dbra{\psi_2}\Bigl(\int\Dketbra{x}{x} \mdiff x\Bigr)\Dket{\psi_1}=\int\Dbraket{\psi_2}{x}\Dbraket{x}{\psi_1}\mdiff x\\
&=\int\psi_2^*(x)\psi_1(x) \mdiff x,
\end{split}
\end{equation*}
as required. The position kets $\Dket{x}$ may be interpreted as eigenstates of a position operator $\xop$, having eigenvalues $x$, such that
\begin{equation}\label{eqA.4}
\xop\Dket{x}=x\Dket{x}\,,
\end{equation}
with $\xop$ having the spectral representation
\begin{equation}\label{eqA.5}
\xop=\int\Dket{x}x\Dbra{x} \mdiff x\\.
\end{equation}
Any $\Dket{\psi}$ can be expressed as a linear superposition of these position eigenstates weighted by the amplitudes $\psi(x)$; \viz,
\begin{equation}\label{eqA.6}
\begin{split}
\Dket{\psi} = \int \mdiff x^\prime\psi(x^\prime)\Dket{x^\prime}\,,
\end{split}
\end{equation}
since then, with \eqref{eqA.2},
\begin{equation}\label{eqA.7}
\begin{split}
\Dbraket{x}{\psi} &= \int \mdiff x^\prime\psi(x^\prime)\Dbraket{x}{x^\prime}
= \int \mdiff x^\prime\psi(x^\prime)\delta(x^\prime-x)\\
&= \psi(x)\,,
\end{split}
\end{equation}
consistent with \eqref{eqA.1}.

For the purpose of representing the Fourier transform $\tilde{\psi}(k)$ of wave function $\psi(x)$, defined here as
\begin{subequations}\label{eqA.8}
\begin{equation}\label{eqA.8a}
\tilde{\psi}(k)=\int\me^{-\mi k x}\psi(x) \mdiff x\,,
\end{equation}
with inverse
\begin{equation}\label{eqA.8b}
\psi(x)=\frac{1}{2\pi}\int\me^{\mi k x}\tilde{\psi}(k) \mdiff k\,,
\end{equation}
\end{subequations}
Dirac also introduces the kets $\Dket{k}$ for the conjugate wave vector $k$ with the property
\begin{equation}\label{eqA.9}
\Dbraket{x}{k}=\me^{\mi k x}\,,	
\end{equation}
with
\begin{equation}\label{eqA.10}
	\Dbraket{k}{k^\prime}=2\pi\delta(k^\prime-k)\,,
\end{equation}
so that
\begin{equation}\label{eqA.11}
	\frac{1}{2\pi}\int\Dketbra{k}{k} \mdiff k = 1\,.
\end{equation}
Then we have the identification
\begin{equation}\label{eqA.12}
	\Dbraket{k}{\psi}=\tilde{\psi}(k)\,,
\end{equation}
consistent with the integral definitions $\tilde{\psi}(k)$ and its inverse $\psi(x)$, and with the requirement,
\begin{equation*}
\begin{split}
\Dbraket{x}{x^\prime} &= \Dbra{x}\Bigl(\int\Dketbra{k}{k}\frac{\mdiff k}{2\pi}\Bigr)\Dket{x^\prime}=\frac{1}{2\pi}\int \me^{\mi k(x-x^\prime)} \mdiff k\\
&=\delta(x^\prime-x).
\end{split}
\end{equation*}
By analogy with $\Dket{x}$, $\Dket{k}$ can be identified as the eigenstate of the wave vector operator $\kop$, such that
\begin{equation}\label{eqA.13}
	\kop\Dket{k}=k\Dket{k}\,,
\end{equation}
with $\kop$ having the spectral representation
\begin{equation}\label{eqA.14}
	\kop=\int\Dket{k}k\Dbra{k} \frac{\mdiff k}{2\pi}\,.
\end{equation}
The operator $\kop$ can be represented in real space---\ie, ``mapped'' onto the $x$ axis---in the following manner. Starting with $\kop\Dket{\psi}$ we form 
\begin{equation*}
\Dbra{x}\kop\Dket{\psi}=\Dbra{x}\kop\bigl(\int\Dketbra{y}{y}dy\bigr)\Dket{\psi}=\int K(x,y)\psi(y)dy\,,
\end{equation*}
where, making use of \eqref{eqA.14},
\begin{equation*}
K(x,y)=\Dbra{x}\kop\Dket{y}=\int\me^{\mi k (x-y)}k\frac{\mdiff k}{2\pi}=\mi\frac{\partial\delta(x-y)}{\partial y}\,.
\end{equation*}
Then integration by parts yields
\begin{equation*}
\int K(x,y)\psi(y)dy = -\mi\int\frac{\partial\psi(y)}{\partial y}\delta(x-y) = -\mi\frac{\partial\psi(x)}{\partial x}\,.
\end{equation*}
Therefore $\kop\Dket{\psi}$ is mapped onto $x$ as $-\mi\partial\psi(x)/\partial x$, allowing us to say that
$\kop$ itself is represented on the $x$--axis by 
\begin{equation}\label{eqA.15}
\kop\rightarrow \pop=-\mi\frac{\partial}{\partial x}\,.
\end{equation}
\end{appendix}

\bibliography{NBERK11918}
\end{document}